\documentclass[preprint,12pt]{aastex}
\usepackage[dvips]{epsfig}

\shorttitle{Horizons}
\shortauthors{Bj\"{o}rnsson and Gudmundsson}

\begin{document}
\singlespace
%\onecolumn
%% LaTeX will automatically break titles if they run longer than
%% one line. However, you may use \\ to force a line break if
%% you desire.

\title{DARK ENERGY AND THE OBSERVABLE UNIVERSE}

%% Use \author, \affil, and the \and command to format
%% author and affiliation information.
%% Note that \email has replaced the old \authoremail command
%% from AASTeX v4.0. You can use \email to mark an email address
%% anywhere in the paper, not just in the front matter.
%% As in the title, you can use \\ to force line breaks.

\author{Einar H. Gudmundsson\altaffilmark{} and Gunnlaugur Bj{\"{o}}rnsson\altaffilmark{}}
\affil{Science Institute, University of Iceland \\ 
Dunhaga 3, IS-107 Reykjavik, Iceland}

%\author{C. D. Biemesderfer\altaffilmark{4,5}}
%\affil{National Optical Astronomy Observatories, Tucson, AZ 85719}
%\email{aastex-help@aas.org}

%\and

%\author{R. J. Hanisch\altaffilmark{5}}
%\affil{Space Telescope Science Institute, Baltimore, MD 21218}

%% Notice that each of these authors has alternate affiliations, which
%% are identified by the \altaffilmark after each name.  Specify alternate
%% affiliation information with \altaffiltext, with one command per each
%% affiliation.

%\altaffiltext{1}{Visiting Astronomer, Cerro Tololo Inter-American Observatory.
%CTIO is operated by AURA, Inc.\ under contract to the National Science
%Foundation.}
%\altaffiltext{2}{Society of Fellows, Harvard University.}
%\altaffiltext{3}{present address: Center for Astrophysics,
%    60 Garden Street, Cambridge, MA 02138}
%\altaffiltext{4}{Visiting Programmer, Space Telescope Science Institute}
%\altaffiltext{5}{Patron, Alonso's Bar and Grill}

%% Mark off your abstract in the ``abstract'' environment. In the manuscript
%% style, abstract will output a Received/Accepted line after the
%% title and affiliation information. No date will appear since the author
%% does not have this information. The dates will be filled in by the
%% editorial office after submission.

\begin{abstract} 
We consider ever-expanding Big Bang models with a cosmological constant, $\Lambda$, 
and investigate in detail the evolution of the observable part of the universe.
We also discuss quintessence models from the same point of view.

A new concept, the $\Lambda$-sphere (or $Q$-sphere, in the case of quintessence) is introduced. 
This is the surface in our visible universe which bounds the region where dark energy dominates 
the expansion, and within which the universe is accelerating. 
We follow the evolution of this surface as the universe expands, and we also
investigate the evolution of the particle and event horizons as well as the 
Hubble surface.  We calculate the extent of the observable universe and the portion of it  
that can be seen at different epochs. Furthermore, we trace the changes in redshift, apparent 
magnitude and apparent size of distant sources through cosmic history. 
Our approach is different from, but complementary to, most other
contemporary investigations, which concentrate on the past light cone at the present epoch. 

When presenting numerical results we use the FRW world model with $\Omega_{m0} = 0.30$ and 
$\Omega_{\Lambda0} = 0.70$ as our standard cosmological model. In this model the 
$\Lambda$-sphere is at a redshift of $0.67$, and within a few Hubble times the event
horizon will be stationary at a fixed proper distance of $5.1$ Gpc (assuming $h_0 = 0.7$).
All cosmological sources with present redshift larger than $1.7$ have by now crossed the
event horizon and are therefore completely out of causal contact.
\end{abstract}

%% Keywords should appear after the \end{abstract} command. The uncommented
%% example has been keyed in ApJ style. See the instructions to authors
%% for the journal to which you are submitting your paper to determine
%% what keyword punctuation is appropriate.


%% From the front matter, we move on to the body of the paper.
%% In the first two sections, notice the use of the natbib \citep
%% and \citet commands to identify citations.  The citations are
%% tied to the reference list via symbolic KEYs. The KEY corresponds
%% to the KEY in the \bibitem in the reference list below. We have
%% chosen the first three characters of the first author's name plus
%% the last two numeral of the year of publication as our KEY for
%% each reference.

%% We may use the \subsection command to set off
%% a subsection.  \footnote is used to insert a footnote to the text.

%\subsection{Formalism} \label{bozomath}
%\footnote{Footnotes can be inserted like this.}

%% Observe the use of the LaTeX \label
%% command after the \subsection to give a symbolic KEY to the
%% subsection for cross-referencing in a \ref command.
%% You can use LaTeX's \ref and \label commands to keep track of
%% cross-references to sections, equations, tables, and figures.
%% That way, if you change the order of any elements, LaTeX will
%% automatically renumber them.

%% This section also includes several of the displayed math environments
%% mentioned in the Author Guide.

%------------------------------ 1 ---------------------------------------
\section{INTRODUCTION}
\label{intro}

At present there is growing evidence that the expansion of our visible universe is accelerating.
Two independent research groups using Type Ia supernovae as standard candles have 
discovered signs of a small 
positive cosmological constant in the Hubble diagram \citep{per99,gar98,rie98,arie01}. 
Further evidence comes from investigations of anisotropies
in the cosmic microwave background as well as large scale structure and the age of the universe 
(for recent reviews and references see \cite{car00}, \cite{sah00} and \cite{bah99}). 

Assuming standard cosmological theory \citep{pea99,pee93} these observations indicate 
that matter, including dark matter, contributes about 30\% of the critical density and 
an effective cosmological constant or dark energy about 70\%.
These results indicate that our universe 
is close to being flat 
and has a Big Bang origin. Furthermore, we are living at a time in cosmic history when the 
cosmological constant, or something that mimics its effects, is already
dominating the expansion.

In this paper we shall investigate the evolution of ever-expanding Big Bang world models with 
a cosmological constant, or a quintessence field, with particular emphasis on the evolution of the 
observable universe, {\it i.e.}\ our past light cone and relevant observables such as redshift of 
cosmic sources, their apparent magnitude and apparent size. 
The properties of the particle horizon, the Hubble surface and the event horizon 
will also be discussed.

We introduce a new concept, the $\Lambda$-sphere (or $Q$-sphere, in the case of quintessence), 
which is the surface in our visible universe that bounds the region where dark energy dominates 
the expansion, and within which the universe is accelerating. We track the evolution of 
this surface through cosmic history.

Our methods are in many ways similar to the ones used in our work on the evolution of 
closed world models without a cosmological constant \citep{bjo95} in which an extensive list of 
references to earlier work on cosmic evolution can be found (see also \cite{ada97} for a  
different perspective and further references). We emphasize that  our approach is
different from, but complementary to, most other recent investigations which concentrate on applying
new and old cosmological tests, such as the $m - z$ relation, to the present light cone.

The paper is organized as follows. We begin in \S \ref{FRW} by reviewing the basic definitions 
and results of standard cosmology that are relevant to our discusson. In \S \ref{obs} 
we discuss the
light cone, the particle and event horizons as well as the Hubble surface for a fundamental observer. 
In \S \ref{evolution}, \ref{changes} and \ref{causal} we present our results for the evolution of 
observable quantities such as redshift, apparent magnitude and apparent angular size
and discuss the properties of the $\Lambda$-sphere as well as the question of causal connections. 
Similar methods are then used in \S \ref{quintessence} 
to investigate the effects of quintessence and 
in \S \ref{discussion} we conclude the paper.

%------------------------------ 2 --------------------------------------
\section{THE WORLD MODELS}
\label{FRW}

The space-time metric of the standard spatially homogeneous and isotropic
Friedmann-Robert\-son-Walker (FRW) world models can be written in the form (see e.g.\ 
\cite{wei72,pee93,pea99})
\begin{equation}
\label{metric}
ds^2 = - c^2 dt^2 + R^2(t)\left(\frac{dr^2}{1 - kr^2} + r^2 d\Omega^2 \right)\;,
\end{equation}
where  $d\Omega^2 = d\theta^2 + \sin^2{\theta}\;d\phi^2$, $(r, \theta, \phi)$ are comoving 
spherical coordinates,  $t$ is the cosmic proper time and $c$ is the velocity of light.
$R = R(t)$ is the universal scale factor and  $k$ the curvature scalar which takes 
one of three possible values according to whether the universe is open ($k = - 1$), flat ($k = 0$)
or closed ($k = + 1$).

The time evolution of the models is determined by the Einstein field equations 
for a universe composed of a perfect fluid and by the equation of energy-momentum conservation.
These equations can be reduced to a system of two differential equations in the form 
\begin{equation}
\label{eq1}
{\dot R}^2 = \frac{8 \pi G}{3} \rho R^2 - k c^2
\end{equation}
and
\begin{equation}
\label{eq2}
 \frac{d \rho}{d R} = - \frac{3}{R} \left(\rho + \frac{P}{c^2}\right) \;.
\end{equation}
Here the dot means differentiation with respect to $t$, 
$G$ is Newton's constant of 
gravitation and 
the total mass-energy density, $\rho$, and the total pressure, $P$, are given by
\begin{equation}
\rho =  \rho_m + \rho_r + \rho_{\Lambda} + \rho_Q
\end{equation}
and 
\begin{equation}
P =   P_m + P_r + P_{\Lambda} + P_Q
\end{equation}
respectively. Above, $\rho_m$ is the mass-energy density of non-relativistic matter, 
including non-relativistic dark matter, 
and $\rho_r$ that of radiation, including all relativistic particles. 
Similarily $P_m$ and $P_r$ are the pressure of matter and radiation respectively.
The term $\rho_{\Lambda}$ is the mass-energy density
associated with Einstein's cosmological constant $\Lambda$,
\begin{equation}
\rho_{\Lambda} = \frac{c^2 \Lambda}{8 \pi G}\;,
\end{equation}
and the corresponding effective pressure is given by
\begin{equation}
\label{lambdaeos}
 P_{\Lambda} = - \rho_{\Lambda} c^2\;.
\end{equation}
The mass-energy density and pressure of quintessence are denoted by $\rho_Q$ and $P_Q$ 
respectively.

Note that in order to solve equations (\ref{eq1}) and (\ref{eq2}) one also needs 
equations of state for matter, radiation and quintessence. In general one can write for each
component
\begin{equation}
\label{eos}
 P_i = w_i \rho_i c^2\;,
\end{equation}
where $i$ stands for each of $m$, $r$, $\Lambda$ or $Q$ with $w_m = 0$,  $w_r = 1/3$, 
$w_{\Lambda} = -1$ and $-1 < w_Q < 0$ (for a further discussion on
the cosmic equation of state
see e.g.\ \cite{gud90} and references therein. 
That paper uses $\alpha_i$ instead of $w_i$ as a pressure parameter, 
with $\alpha_i = 3 w_i$). 

In this paper we shall primarily be concerned with the effects of a cosmological constant 
on the evolution of the observable 
universe with emphasis on cosmic epochs in which matter dominates over 
radiation. The effects of quintessence, which  could
mimic the effects of a true 
cosmological constant at the present epoch, will be considered in \S~\ref{quintessence}. 
In later sections we shall  assume that $\rho_r = 0$ and $P_r = 0$. 

The total mass-energy density parameter is defined by
\begin{equation}
\Omega = \sum \Omega_i\;,
\end{equation}
where
$\Omega_i = \rho_i / \rho_c$
is the contribution of each component, $m$, $r$, $\Lambda$ and $Q$, and
$\rho_c = {3 H^2}/{8 \pi G} = 1.9 \times 10^{-29} h^2 \;{\rm g/cm^3}$
is the critical density.
Here $h = H /(100~{\rm km s}^{-1}{\rm Mpc^{-1}})$, where  $H = {\dot R}/R$ is the Hubble constant.
Note that since $H$ is a function of time, so is
$\rho_c$ and hence  $\Omega$ (for $k = \pm 1$).
The relation between $\Omega$ and other
cosmological parameters can be found by rewriting equation~(\ref{eq1}) in the form
\begin{equation}
\label{omegaeq}
\Omega = 1 + \frac{k c^2}{R^2 H^2} =  1 + \frac{k c^2}{{\dot R}^2}\;.
\end{equation}

Using equation~(\ref{omegaeq}) and the fact that after decoupling each of the cosmic components 
$m$, $r$, $\Lambda$ and $Q$ seperately satisfy equation~(\ref{eq2}),
the dynamical equation~(\ref{eq1}) can be written in a convenient form as
\begin{equation}
\label{main}
\frac{d a}{d \tau} = \left\{\Omega_{m0} \left(\frac{1}{a} - 1 \right) 
                         + \Omega_{r0} \left(\frac{1}{a^2} - 1 \right) 
                         + \Omega_{\Lambda 0} \left(a^2 - 1\right) 
                      + \Omega_{Q0}\left(\frac{1}{a^{1 + 3w_Q}} - 1 \right) + 1\right\}^{1/2}
\end{equation}
for ever-expanding models.
Here $w_Q$ is assumed to be constant, $\tau = t / t_{H_0}$ is a new dimensionless
cosmic time variable defined in terms of the Hubble time at the present epoch,
$t_{H_0} = 1/H_0 = 9.8 h_{0}^{-1}$ Gyr, and $a = R/R_0$ is the scale factor normalized 
to its present value.
In equation~(\ref{main}) 
and in what follows we
denote the values of quantities at the present time by the subscript $0$. Note 
that $a(\tau_0) = ({d a}/{d \tau})_{\tau = \tau_0} = 1$ 
and $H = {\dot R}/{R} = H_0 ({1}/{a})({d a}/{d \tau})$.
For future reference we also remind the reader of the definition of the Hubble radius, $R_H = c/H$. 
Its present day value is $R_{H0} = c/H_0 = 3.0 h_0^{-1}~{\rm Gpc}$, where 
$h_0 = H_0 /(100~{\rm km s}^{-1}{\rm Mpc^{-1}})$.

In order to determine $a$ as a function of 
$\tau$ one simply integrates equation~(\ref{main}) for given values of $\Omega_{m0}$, $\Omega_{r0}$,
$\Omega_{\Lambda 0}$ and $\Omega_{Q0}$ as well as $w_Q$.  
Note that the age of the universe at the present epoch is given by
\begin{equation}
\label{age}
\tau_0 = \int_{0}^{1} \left(\frac{d a}{d \tau}\right)^{-1} da  \;.
\end{equation}
 
In the following sections we shall be concerned with the evolution of the subclass of 
FRW models that have a Big Bang origin and continue to expand forever, 
since recent observations indicate that we live in such a universe. For the relevant region
in the $(\Omega_m, \Omega_{\Lambda})$ plane we refer the reader to
equations 11 and 12 and Figure\ 1 in \cite{car92} and   
Figure 7 in \cite{per99}.

%------------------------------- 3 ---------------------------------------
\section{THE OBSERVABLE UNIVERSE}
\label{obs}

In this section we shall discuss various concepts which are necessary 
for an understanding of our observable universe. For this purpose it
is convenient to start by reminding the reader of the definition of
the conformal time $\eta$:  
\begin{equation}
d\eta = \frac{c dt}{R(t)} \;.
\end{equation}
In a Big Bang universe the relation between $\eta$ and
the time variable $\tau$ is therefore given by
\begin{equation}
\label{etatau}
\eta =  \frac{R_{H0}}{R_0} \int_{0}^{\tau} \frac{d \tau}{a (\tau)}\;.
\end{equation}

The comoving conformal radial distance, $\chi$, is defined by
\begin{equation}
d\chi^2 = \frac{dr^2}{1 - kr^2} \;.
\end{equation}
Assuming that we are in the position of a fundamental observer 
at the origin, $\chi =0$ (corresponding to $r = 0$),
we therefore have that
\begin{eqnarray}
\label{rchi}
\chi & = & 
\left\{
\begin{array}{cc}
\sinh^{-1}{(r)} & k = - 1\\
 r              & k = 0\\
\sin^{-1}{(r)}  & k = + 1 \;. 
\end{array} 
\right.
\end{eqnarray}

For our distance calculations we only need the radial coordinate. Setting 
$d\theta = d\phi = 0$ in equation~(\ref{metric}) and using conformal representation, 
the space-time metric becomes very simple: 
\begin{equation}
\label{cmetric}
ds^2 = R^2(\eta)[d\chi^2 - d\eta^2]\;.
\end{equation}

\begin{figure}[t]
\epsscale{1.}
\plottwo{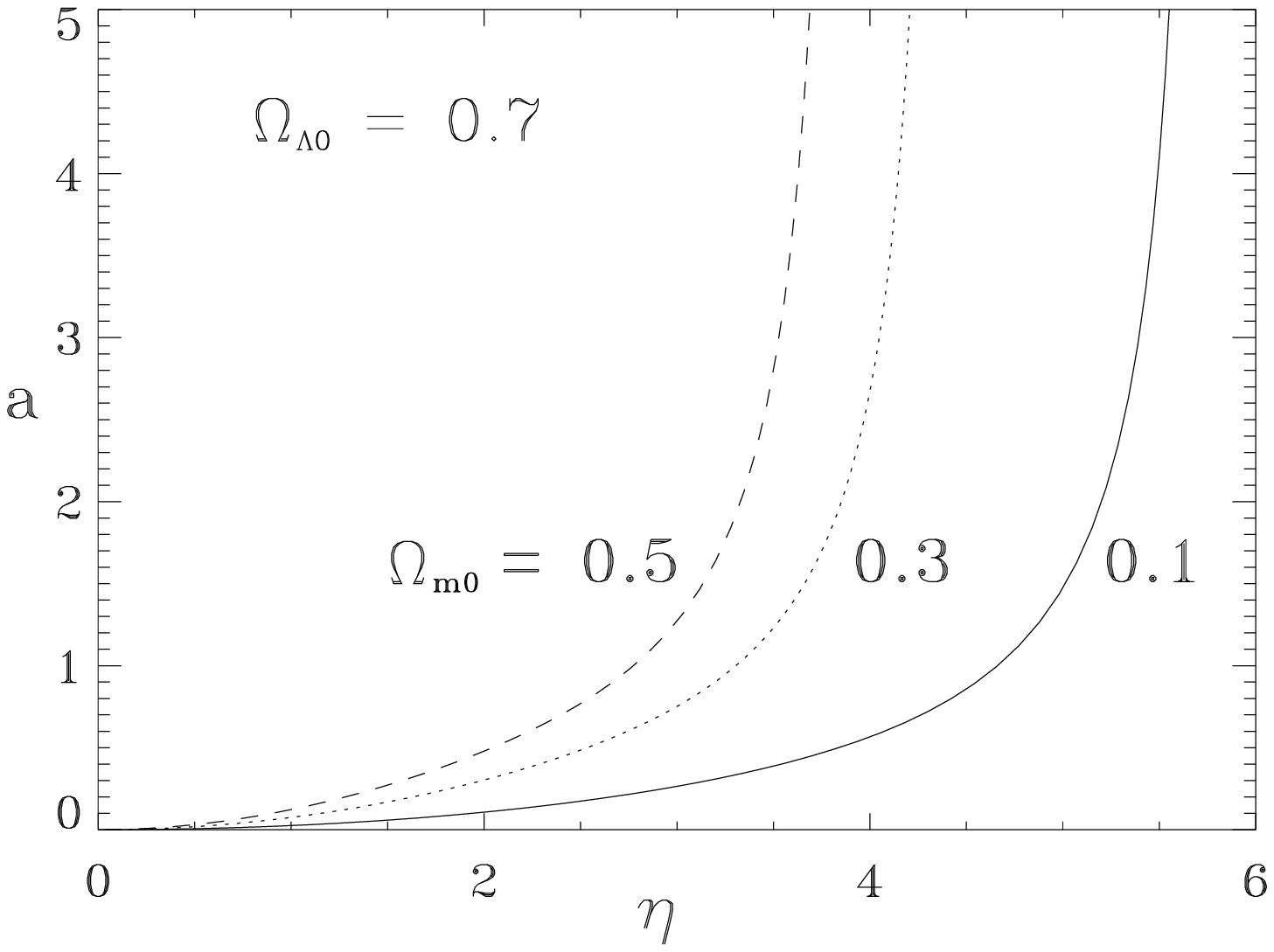}{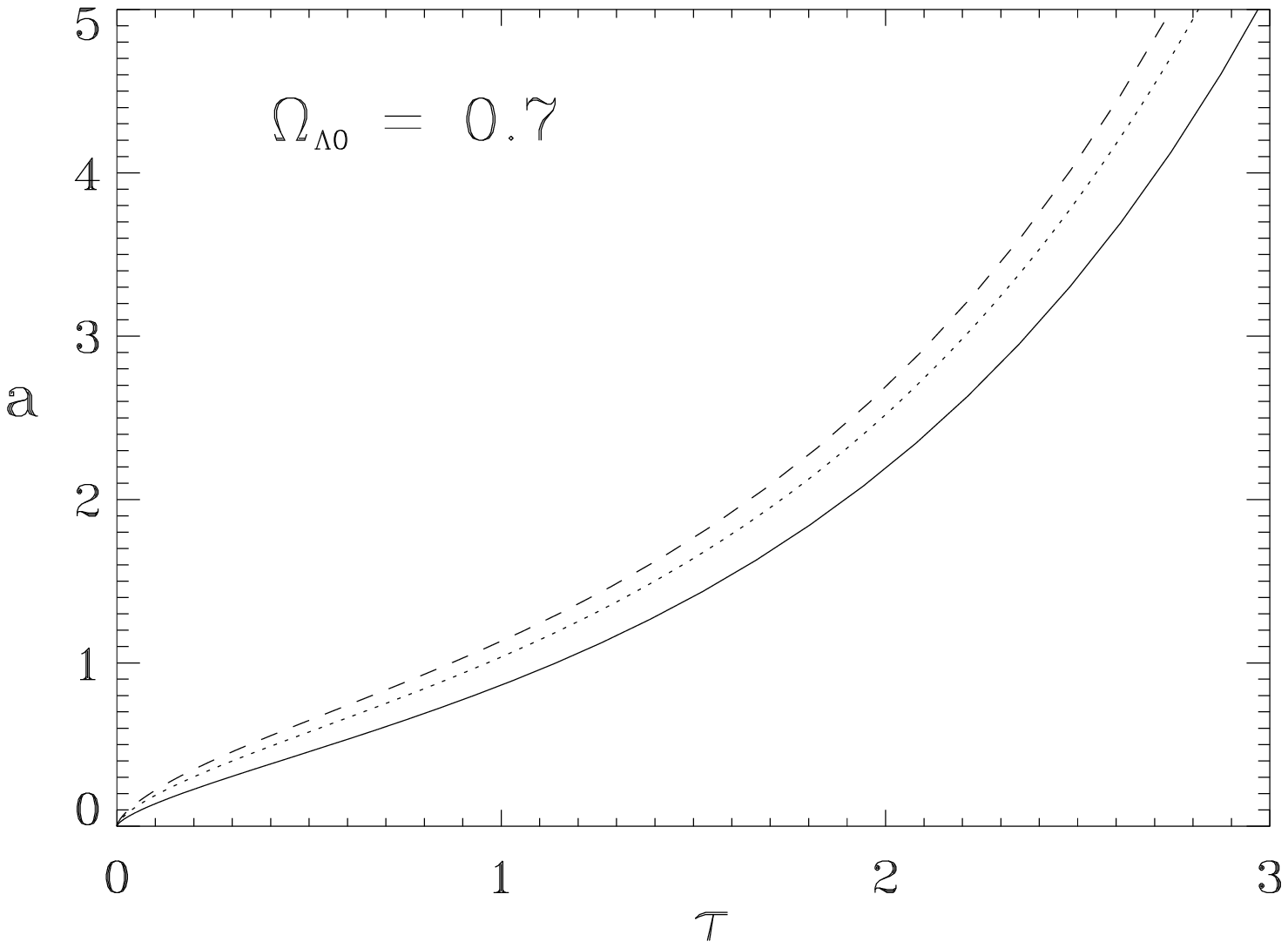}
\epsscale{2.23}
\plottwo{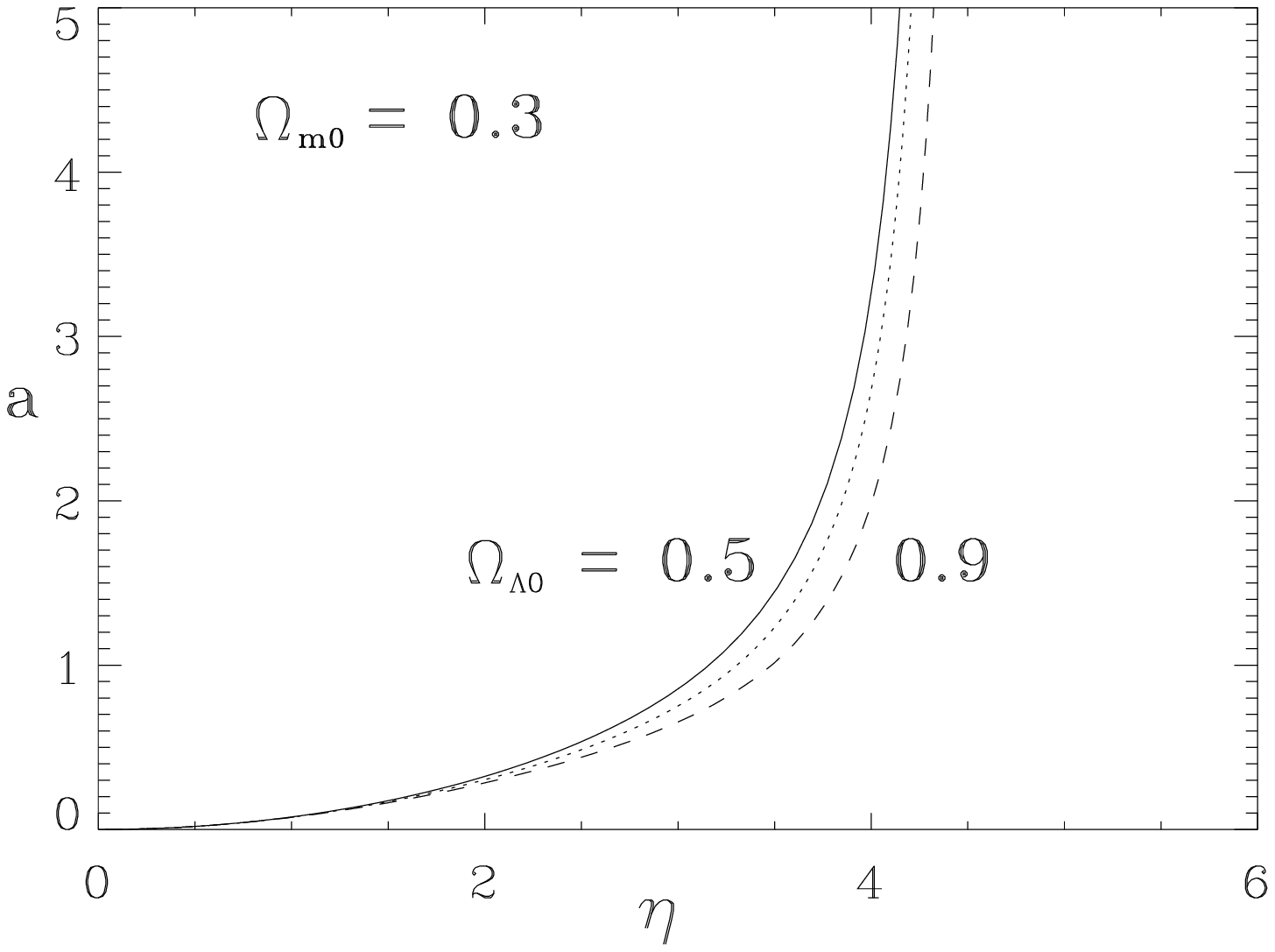}{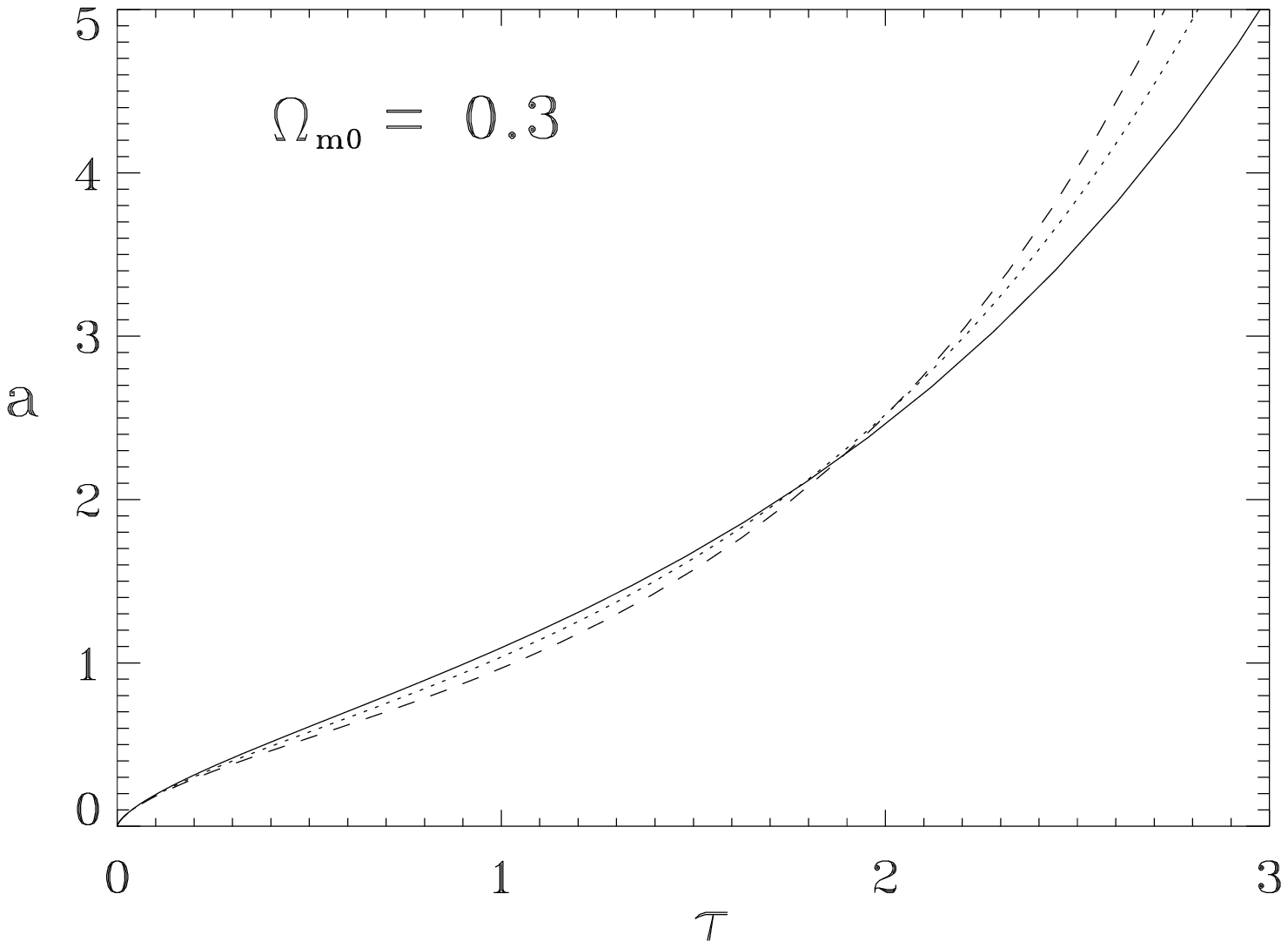}
\caption{(a) The scale factor $a$ as a function of $\eta$ for $\Omega_{\Lambda0}= 0.70$
and  $\Omega_{m0} = 0.10$ (solid curve), $0.30$ (dotted curve) and $0.50$ (dashed curve).
Here and in all figures that follow, $\eta$ is in units of $R_{H0}/R_0$. 
(b) $a$ as a function of $\tau$ for the same values of $\Omega_{\Lambda0}$ and
$\Omega_{m0}$ as in a.
(c) The scale factor as a function of $\eta$ for $\Omega_{m0} = 0.30$ and
$\Omega_{\Lambda0}= 0.50$  (solid curve), $0.70$ (dotted curve) and $0.90$ (dashed curve).
(d) $a$ as a function of $\tau$ for the same values of $\Omega_{\Lambda0}$ and
$\Omega_{m0}$ as in c.  \label{fig1}
}
\end{figure}

Figure~\ref{fig1} shows the scale factor, $a$, both as a function of $\eta$ 
in units of $R_{H0}/R_0$, and $\tau$
for selected values of $\Omega_{m0}$ and $\Omega_{\Lambda 0}$ 
with $\Omega_{r0} = \Omega_{Q0} = 0$. Note in particular that while $\tau$ covers the whole 
range from $0$ to $\infty$, $\eta$ has a maximum value, which we denote by $\eta_{max}$.
A detailed discussion of $\eta_{max}$ will be given in \S~\ref{eventhor}.
In what follows, in the main text as well as in figure captions,
both $\eta$ and $\chi$ will be presented in units of $R_{H0}/R_0$.

%%%%%%%%%%%%%%%%%%%%%%%%%%%%%%%%%%%%%%%%%%%%%%%%
%%%%%%%%%%%%%%%%%%%%%%%%%%%%%%%%%%%%%%%%%%%%%%%

We now discuss in turn the light cone of a fundamental observer, 
his Hubble sphere and his event horizon.

%----------------------------------------
\subsection{The Light Cone}
\label{cones}

The light cone of a fundamental observer
is determined by $ds = 0$, i.e.\ by
\begin{equation}
\label{dlcone}
d\eta =  \pm d\chi \;,
\end{equation}
where we have used equation~(\ref{cmetric}).
Here $+$ corresponds to the future light cone and $-$ to the past light cone. 

Assuming that at time $\eta_0$ we receive a signal emitted at time 
$\eta$  from a source at 
$\chi$, integration of (\ref{dlcone}) gives the equation of our past light cone 
at $\eta_0$ as
\begin{equation}
\label{plcone}
\eta = \eta_0 - \chi \;.
\end{equation}
Similarily our future light cone at $\eta_0$ is given by
\begin{equation}
\label{flcone}
\eta = \eta_0 + \chi \;,
\end{equation}
where $\eta$ is now the time in the future at which an observer at $\chi$ receives
a signal emitted by us at $\eta_0$. The future light cone at the Big Bang 
is sometimes called ``the creation light cone'' \citep{rin56}.

%%%%%%%%%%%%%%%%%%%%%%%%%%%%%%%%%%%%%%%%%%
\begin{figure}[t]
\epsscale{1.}
\plottwo{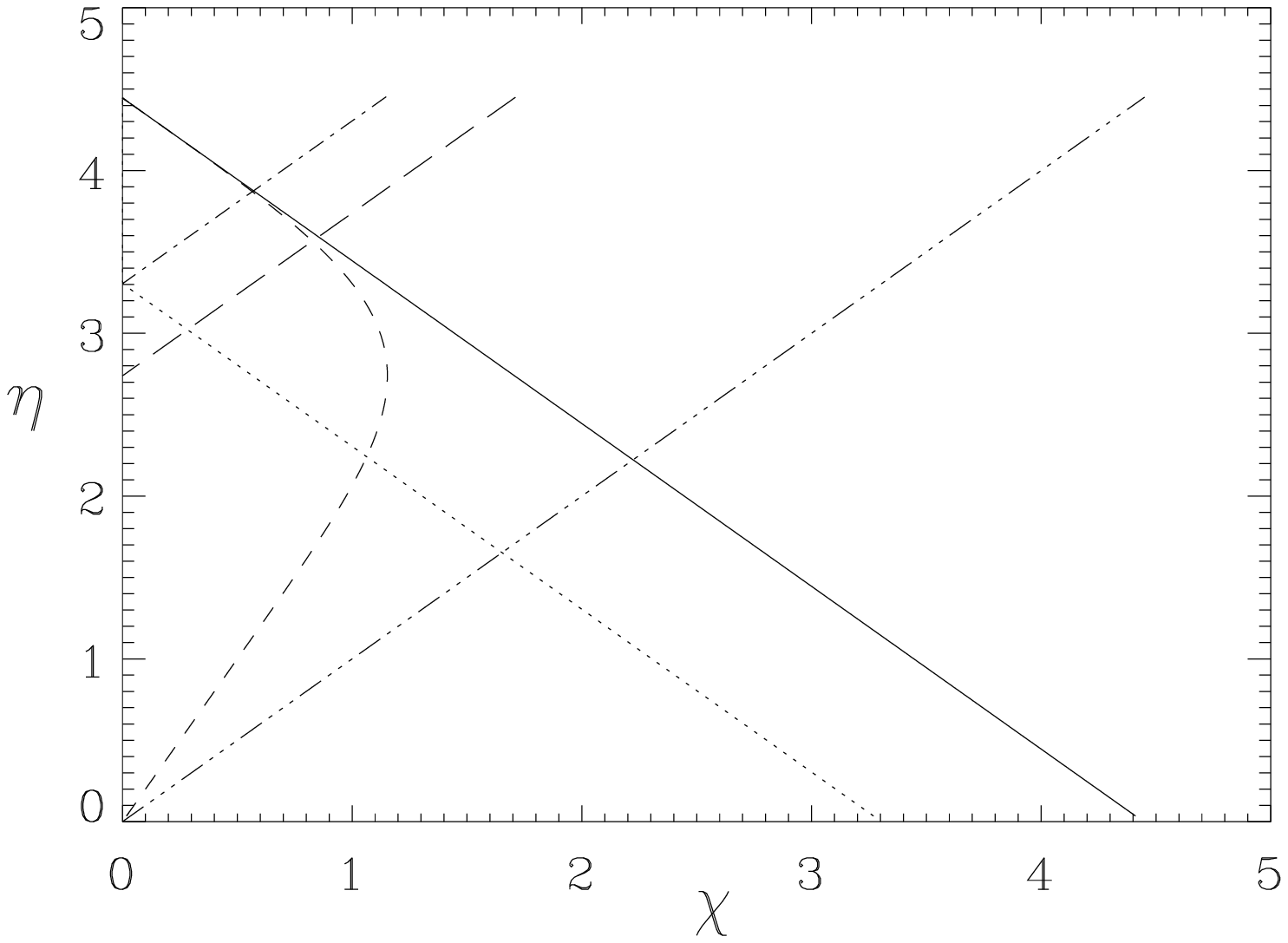}{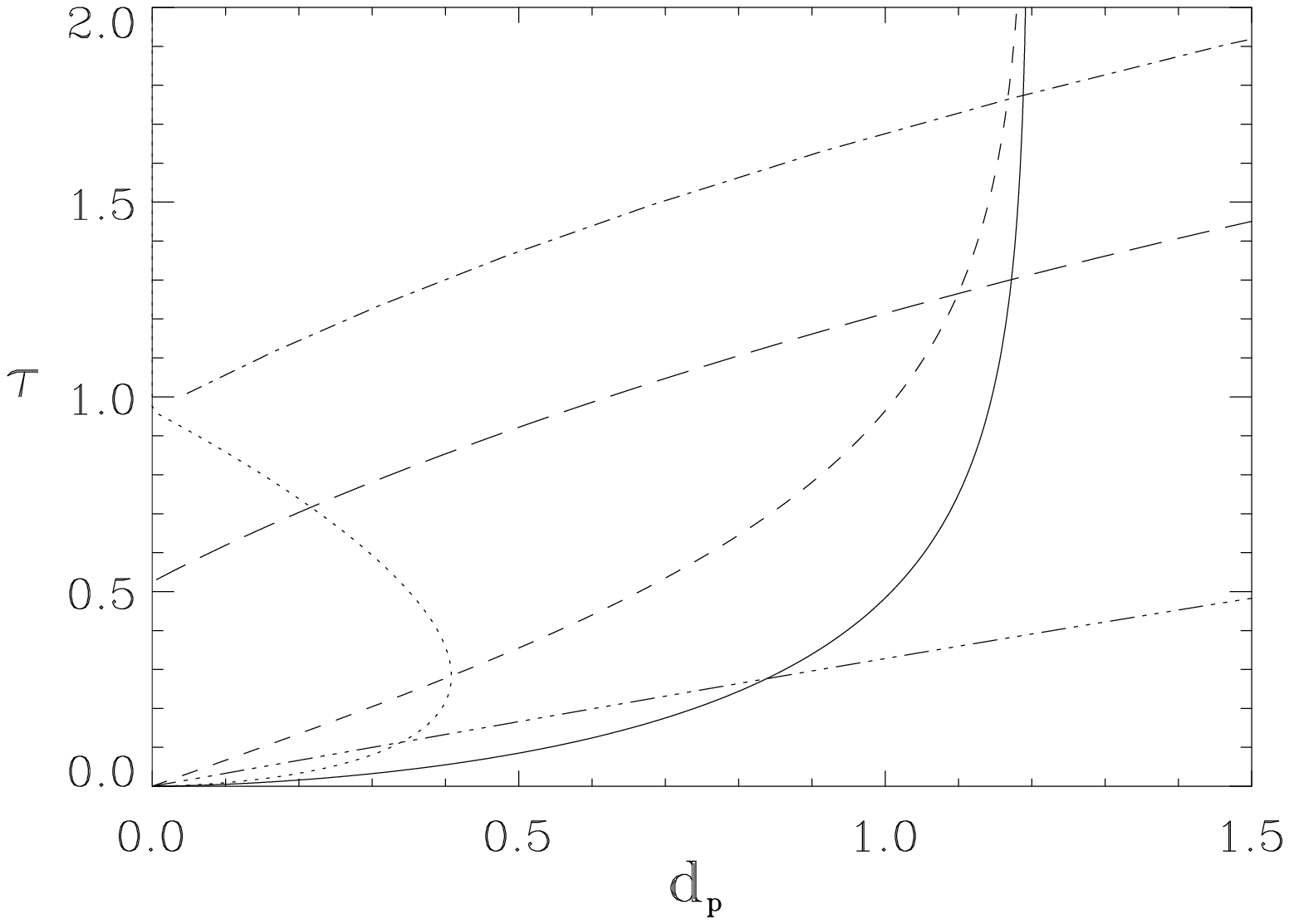}
\caption{(a) Our past light cone (dotted line) and the future light cone (dot-dashed line) 
at the present epoch in a $\eta$ -  $\chi$ diagram for our standard $\Lambda$-model with
$\Omega_{\Lambda0} = 0.70$, $\Omega_{m0} = 0.30$ and $\Omega_{Q0} = \Omega_{r0} = 0$.
Also shown are the Hubble surface (dashed curve), 
the particle or visual horizon (triple dot-dashed line), the event horizon (solid line)
and the $\Lambda$-sphere (long dashed line). 
Here ${{\eta}}_0 = 3.3$ and ${{\eta}}_{max} = 4.5$. 
Note that the creation light 
cone lies right on top of the particle horizon (visual horizon).
(b) Same curves  in a $\tau$ - $d_p$ diagram, where $d_p =  d/R_{H0}$ and $d$ is the 
proper distance. Note that $\tau_0 = 0.96$, and therefore $t_0 = 13.5$ Gyrs 
if $h_0 = 0.70$. \label{fig2}
}
\end{figure}
%%%%%%%%%%%%%%%%%%%%%%%%%%%%%%%%%%%%%%%%%%%

Figure~\ref{fig2} shows our past and future light cones in a model with
$\Omega_{m0} = 0.30$, $\Omega_{\Lambda0} = 0.70$ and $\Omega_{Q0} = \Omega_{r0} = 0$. 
In what follows, 
this will be our standard cosmological model when presenting numerical results,
and we shall refer to it as our standard $\Lambda$-model. 
In this model $\tau_0 = 0.96$ which corresponds to $\eta_0 = 3.3$,
whereas the end of time $\tau = \infty$ corresponds to $\eta = \eta_{max} = 4.5$
(our numerical results will generally be given with two significant figures).
The distance $d_p$ is the normalized proper distance
$d_p = d / R_{H0}$, with the proper distance given by $d = d(\tau) = R(\tau)\chi$.
Figure~\ref{fig2} also shows the creation light cone as well as  the Hubble surface, 
the particle horizon, the event horizon and the $\Lambda$-sphere
which will all be discussed in detail in 
sections \ref{parthor} - \ref{eventhor} and \ref{redbrightsize} below. 

In terms of the time varible $\tau$ and the proper distance, $d = R\chi$, the
past light cone (lc) at $\tau_0$ is given by
\begin{equation}
d_{lc}(\tau) =  R_{H0}\;a(\tau) \left({\int_{0}^{\tau_0}\frac{d\tau}{a(\tau)}} 
                                      - {\int_{0}^{\tau}\frac{d\tau}{a(\tau)}}\right)\;,
\end{equation}
with $0 \leq \tau \leq \tau_0$.

%-----------------------------------------------
\subsection{The Particle Horizon}  
\label{parthor}

At time $\eta$ our particle horizon (ph) is situated at $\chi = \eta$ 
(\cite{rin56}). The proper distance to this horizon is therefore
\begin{equation} 
d_{ph}(\eta) = R\eta\;,
\end{equation}
and as a function of $\tau$ it is given by
\begin{equation} 
d_{ph}(\tau) = R\eta = R_{H0}\;a(\tau) \int_{0}^{\tau} \frac{d\tau}{a(\tau)}\;.
\end{equation}
The horizon is moving away from us at speed 
\begin{equation}
\label{vph}
v_{ph} = \frac{d}{d t}(d_{ph}) = c + H\;d_{ph}(\eta) 
       = c \left(1 + \frac{\eta}{R_{H}/R} \right)\;. 
\end{equation}
Note that comoving sources momentarily at the particle horizon are moving away from us 
at speed $H d_{ph}$. In a universe with a Big Bang beginning we ``see'' these sources as they 
were at $\tau = 0$ and with infinite redshift. 

In our numerical calculations we assume a universe with ordinary
matter and a cosmological constant (or quintessence), i.e.\ a universe with $\Omega_{r0} = 0$.
This basically means that we ignore the expansion dynamics of the early universe. 
For our purposes this is a good approximation. However it should be kept in mind
that the early universe probably went through an inflationary period. With inflation the 
real particle horizon is much further away than the particle horizon obtained by assuming
a dust universe with a cosmological constant (see e.g.~\cite{har91} and references therein). 
The particle horizon presented in our calculations is therefore approximately equal to 
the particle horizon looking back to the cosmic microwave background. This horizon is sometimes
called the ``visual horizon'' \citep{ell93}.
For our standard $\Lambda$-model we find that $d_{ph}(\tau_0) = 3.3 R_{H0}$
and $v_{ph}(\tau_0) = 4.3c$.

%----------------------------------------
\subsection{The Hubble Surface}
\label{Hubbles}

``The Hubble surface'' (hs; \cite{har91}) is the instantaneous set of points  
which at time $\eta$  are moving away from us at the speed of light. 
Their proper distance is given by the velocity-distance law as 
\begin{equation}
d_{hs}(\eta) = \frac{c}{H} = R_{H} \;, 
\end{equation}
and hence the conformal distance is
\begin{equation}
\chi_{hs}(\eta) = \frac{d_{hs}(\eta)}{R}\;. 
\end{equation}
In terms of  $\tau$, the proper distance to the Hubble surface is given by
\begin{equation} 
d_{hs}(\tau) = R \chi_{hs} = R_{H0}\;a(\tau)  \left(\frac{d a}{d\tau}\right)^{-1}
\end{equation}
and it is moving away from us at speed
\begin{equation}
v_{hs} = \frac{d}{d t}(d_{hs}) = c(1 + q)\;, 
\end{equation}
where
$q = - R \ddot{R} / {\dot{R}}^2 = - a ({d^2 a}/{d \tau^2})/({d a}/{d \tau})^2$  
is the deceleration parameter. 
Note that if the cosmological constant dominates the expansion then
$q = - 1$ and $v_{hs} = 0$.
For our standard $\Lambda$-model we find that $q_0 = -0.85$ and hence the present speed of the 
Hubble surface is $v_{hs}(\tau_0) = 0.15 c$.

In Figure~\ref{fig2} we show the evolution of the Hubble surface by the dashed curve. Note
that at a given time sources beyond the Hubble surface are moving away 
from us with speed greater than $c$,
whereas sources inside the surface are receeding with speed less than $c$. 
From the figure one can also see that on a 
cosmic timescale the Hubble surface rapidly approaches the event horizon and 
within only a few Hubble times after the Big Bang the two
have practically merged, never to part again.

%-------------------------------
\subsection{The Event Horizon}
\label{eventhor}

If the world line of a source at conformal distance $\chi$ 
intersects our past light cone at time $\eta_0$ then we see it as it was at
$\eta = \eta(\tau)$ where
\begin{equation}
\chi = \eta_0 - \eta = \int_{\eta}^{\eta_0}d\eta 
     = \frac{R_{H0}}{R_0} \int_{\tau}^{\tau_0} \frac{d \tau}{a(\tau)}\;. 
\end{equation}
 Now suppose that for an ever-expanding world model 
the integral $\int_{\tau}^{\infty} a^{-1}d\tau$ is  finite.
Then the conformal time 
\begin{equation}
\label{etamax1}
\eta_{max} =  \frac{R_{H0}}{R_0} \int_{0}^{\infty}\frac{d \tau}{a (\tau)}
\end{equation}
corresponding to $\tau_0 = \infty$ is finite
and the conformal distance $\chi_{eh}$ given by
\begin{equation}
\label{eventh}
\chi_{eh} = \eta_{max} - \eta = \int_{\eta}^{\eta_{max}}d\eta
      = \frac{R_{H0}}{R_0} \int_{\tau}^{\infty} \frac{d \tau}{a(\tau)}
     = \frac{R_{H0}}{R_0} \left(\int_{0}^{\infty} \frac{d \tau}{a(\tau)} 
        - \int_{0}^{\tau} \frac{d \tau}{a(\tau)}\right)\;, 
\end{equation}
is the finite distance to our event horizon (eh) at time $\eta$ (or $\tau$). 
This is because the event horizon is our 
final or ultimate light cone \citep{rin56}. It is defined by
equation~(\ref{eventh}) and shown for our standard model by 
the solid curve in Figure~\ref{fig2}. 
All events on the event horizon will first be ``seen'' 
by us at the end of time ($\eta = \eta_{max}$ corresponding to 
$\tau = \infty$) and with infinite redshift (see \S \ref{evolution}). 
Events beyond this horizon will never be seen by us.

The proper distance to the event horizon at time $\tau$ is given by
\begin{equation}
\label{deh}
d_{eh}(\tau) 
   = R \chi_{eh} = R_{H0}\;a(\tau) \left(\int_{0}^{\infty} \frac{d \tau}{a(\tau)} 
        - \int_{0}^{\tau} \frac{d \tau}{a(\tau)}\right)\;, 
\end{equation}
and it is moving away from us with speed
\begin{equation}
\label{veh}
v_{eh} = \frac{d}{d t}(d_{eh}) 
       = c\left(\frac{\eta_{max}}{R_{H0}/ R_0}\right)
         \left(\frac{d a}{ d \tau}\right)_{\tau} - v_{ph} \;,
\end{equation}
where $v_{ph}$ is given by equation~(\ref{vph}).
For our standard $\Lambda$-model we have that $v_{eh}(\tau_0) = 0.14 c$.

From Figure~\ref{fig2}  we see that for the ever-expanding Big Bang models with a cosmological
constant, the event horizon is stationary at a particular proper distance after a 
certain time. Furthermore  the Hubble surface approaches the 
event horizon quite rapidly. For the standard $\Lambda$-model in  Figure~\ref{fig2}, 
we have that the proper distance to the Hubble surface and the
event horizon is fixed at $\approx 1.2 R_{H0}$ at late cosmic epochs. 
For comparison with this ultimate value, we remind the reader that the 
present day value of $d_{hs}$ is $1.0 R_{H0}$, and from equation~(\ref{deh}) we see that 
$d_{eh}(\tau_0) = 1.1 R_{H0}$.

The reason for this limiting behaviour can be  understood in the following way. 
For an ever-expanding model the scale factor increases without limit. After a certain time, 
say $\tau_\bullet$, we see from equation~(\ref{main}) that the $\Lambda$ term dominates completely
and hence it is a good aproximation to 
write 
\begin{equation}
\label{alambda}
a(\tau) =  a(\tau_\bullet)\; e^{\Omega_{\Lambda0}^{1/2}(\tau - \tau_\bullet)}\;, 
          \;\;\;\;  \tau > \tau_\bullet \;.
\end{equation}
From this it follows that 
\begin{equation}
\label{hseh}
d_{eh}(\tau) = d_{hs}(\tau) = R_{H0}\;a(\tau)\left(\frac{d a}{d \tau} \right)^{-1} 
 =\; \frac{R_{H0}}{\Omega_{\Lambda0}^{1/2}}\;, \;\;\;\;  \tau > \tau_\bullet \;. 
\end{equation}
Note that although the proper distance to the event horizon is finite,
its luminosity distance, $d_L$, is infinite. 
However, as will be discussed in more detail in \S \ref{evolution},  
the angular diameter distance, $d_A$, is finite.

For later purposes we also express equation~(\ref{alambda}) in terms of conformal time.
For $\eta$ in the range $\eta_\bullet < \eta \leq \eta_{max}$, 
where $\eta_\bullet$ corresponds to $\tau_\bullet$, we find from
equations~(\ref{etatau}) and~(\ref{alambda}) that
\begin{equation}
\eta  \approx \eta_{max} -  \frac{R_{H0} / R_0}{\Omega_{\Lambda0}^{1/2} a(\eta)}\;,
 \;\;\;\;  \eta_\bullet < \eta \leq \eta_{max}\;,
\end{equation}
and hence
\begin{equation}
\label{aetamax}
a(\eta) \approx   \frac{R_{H0} / R_0}{\Omega_{\Lambda0}^{1/2} (\eta_{max} - \eta)}\;,
 \;\;\;\;  \eta_\bullet < \eta \leq \eta_{max}\;.
\end{equation}

Finally, we remind the reader that only universes with finite $\eta_{max}$ have event horizons. 
For example if the scale factor grows as a power of time, i.e.\ $a \propto \tau^n$, 
then $\eta_{max}$
is finite only if $n > 1$. Models with scale factors growing more slowly than this
have no event horizons,  e.g.\ the open or flat FRW-universes 
with $\Lambda = 0$, and quintessence models
with $- 1/3 < w_Q < 0$ (see also \S~\ref{quintessence}).

%----------------------------- 4 ------------------------------------------
\section{EVOLUTION OF OBSERVABLE QUANTITIES}
\label{evolution}

In this section we shall discuss the time evolution of various observational quantities 
in ever-expanding Big Bang models with a cosmological constant. 

%%%%%%%%%%%%%%%%%%%%%%%%%%%%%%%%%%%%%%%%%%%
\begin{figure}[t]
\epsscale{1.}
\plottwo{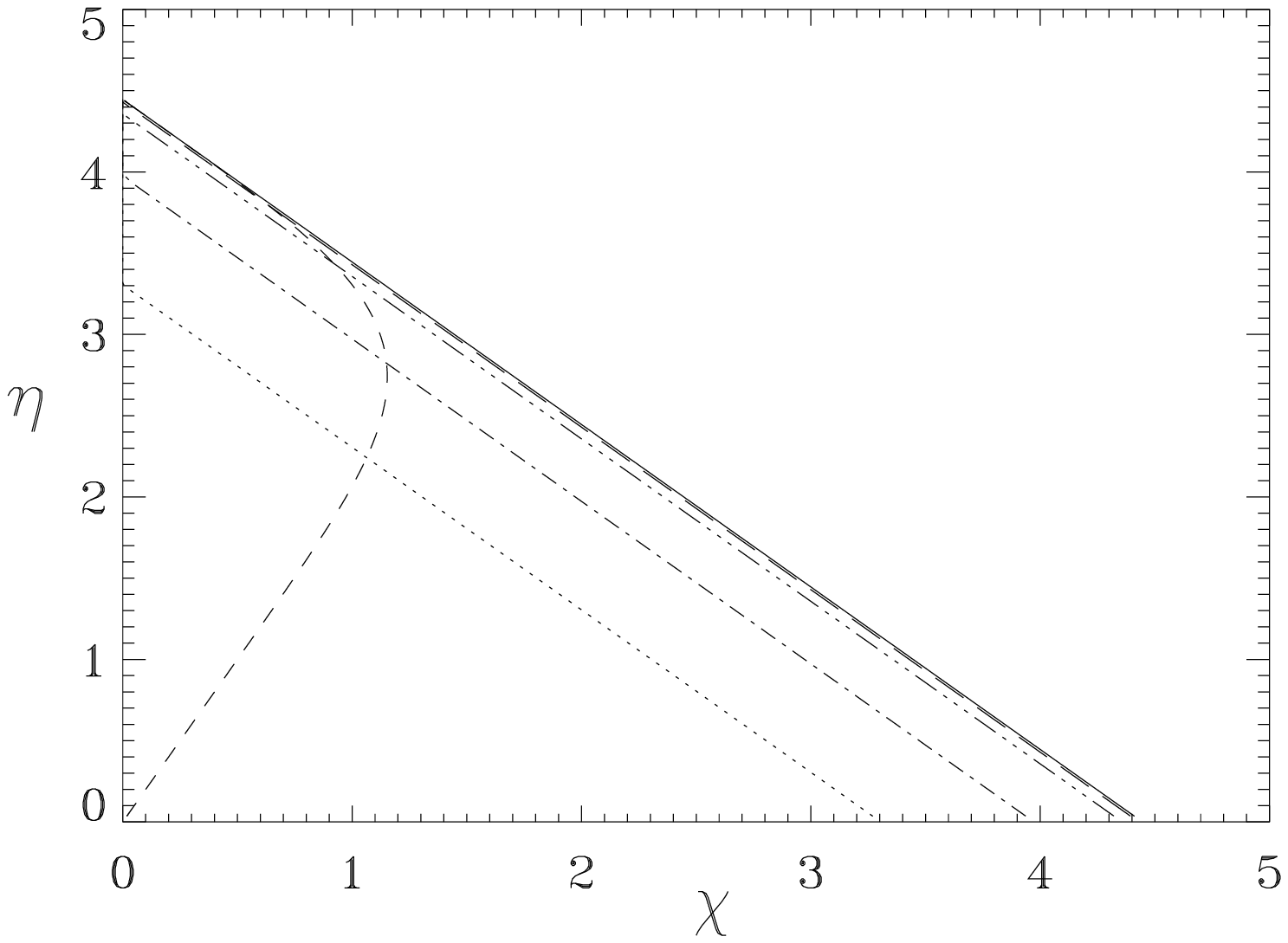}{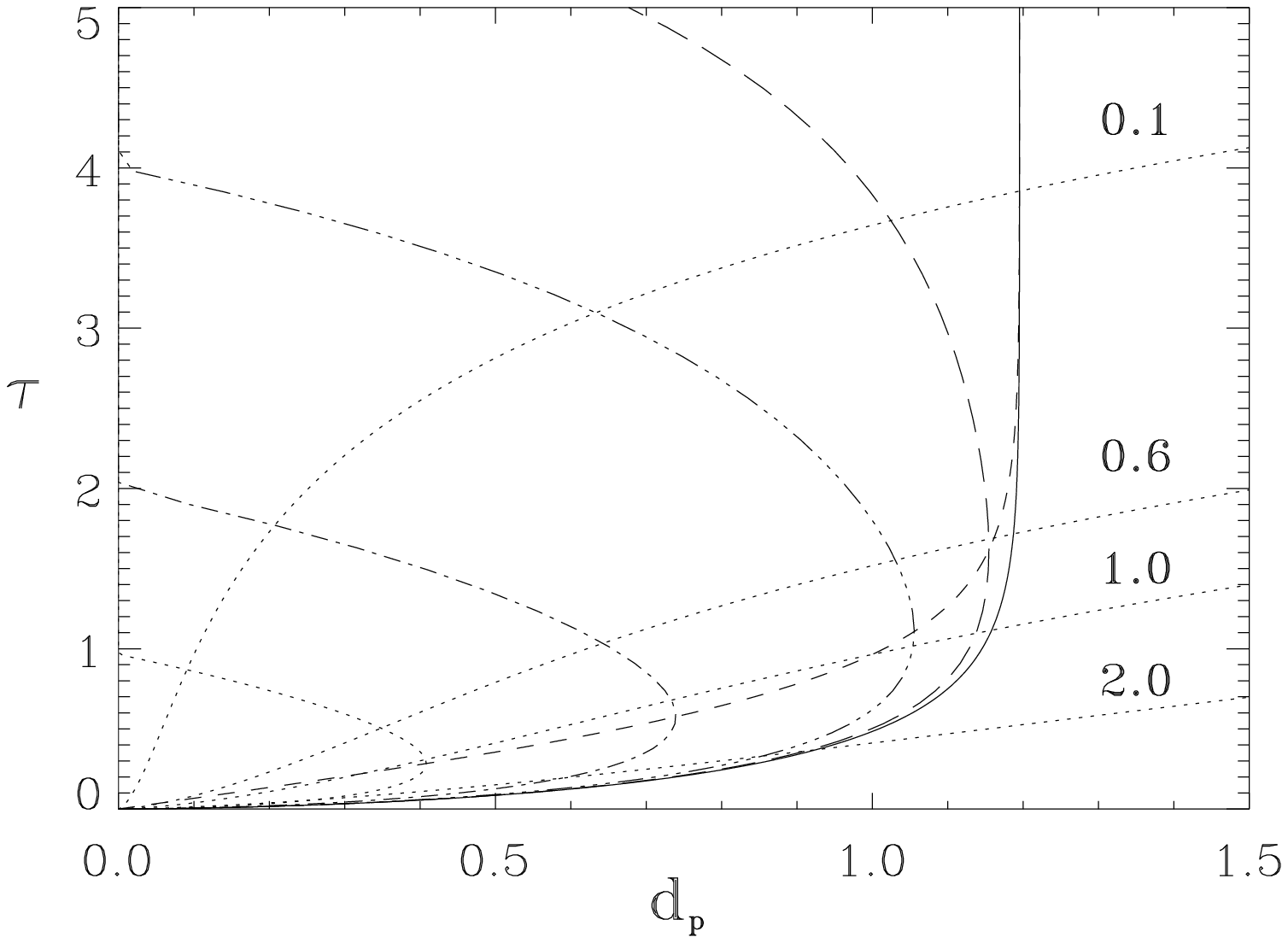}
\caption{(a) The past light cone at various cosmic epochs in a $\eta$ - $\chi$ diagram. 
The world model is the same as in Figure~\ref{fig2}. Also shown are
the event horizon (solid line) and the Hubble surface (short dashed curve).    
(b) Same as a, but in a $\tau$ - $d_p$ diagram. 
In addition, world lines of several sources are also shown as dotted curves, marked by 
the positions of the sources at 
$\chi = 0.10, 0.60, 1.0$ and $2.0$. \label{fig3}
}
\end{figure}
%%%%%%%%%%%%%%%%%%%%%%%%%%%%%%%%%%%%%%%%%%%%%

Consider first 
our past light cone. In Figure~\ref{fig3} we show how it evolves with cosmic time
in our standard $\Lambda$- model. 
As time advances for the observer, his light cone gets closer and closer to the event horizon,
demonstrating that the event horizon corresponds to his final light cone.
One can clearly see how his observable part of the universe is enclosed for all time
within a finite proper volume  with proper radius 
$R_{H0} / \Omega_{\Lambda0}^{1/2} \approx 1.2 R_{H0}$.
As expected, Figure~\ref{fig3}b also shows that 
the Hubble surface (the short dashed curve) crosses the observer's light cones at their maximum 
proper distance from the time axis.
 
The world lines of sources at several different conformal distances, 
$\chi = 0.10, 0.60, 1.0$ and $2.0$, 
are also shown in Figure~\ref{fig3}b (dotted curves). All sources taking
part in the cosmic expansion leave our observable part of the universe in a finite proper time. 
However, just as an observer at rest far away from a black hole never sees infalling objects 
pass the event horizon of the black hole, we shall never see the sources pass through our cosmic
event horizon, although they will rapidly fade away once the cosmological constant 
dominates the expansion.

%-------------------------------------
\subsection{The Redshift and the $\Lambda$-sphere}
\label{redbrightsize}

Assume that a given cosmic  source is located at comoving conformal
distance $\chi$. Its redshift, $z$, as observed at time $\eta_{obs} = \eta(\tau_{obs})$ 
is given by
\begin{equation}
\label{red}
1 + z = \frac{R(\eta_{obs})}{R(\eta_{em})} = \frac{a(\tau_{obs})}{a(\tau_{em})}\;,
\end{equation}
where $\eta_{em} = \eta(\tau_{em})$ is the time at emission, {\it i.e.}\ 
the time at which the world line of the
source crosses the observer's past light cone as it is at 
the time of observation. 

%%%%%%%%%%%%%%%%%%%%%%%%%%%%%%%%%%%%%%%%%%%%%%%
\begin{figure}[t]
\epsscale{1.}
\plottwo{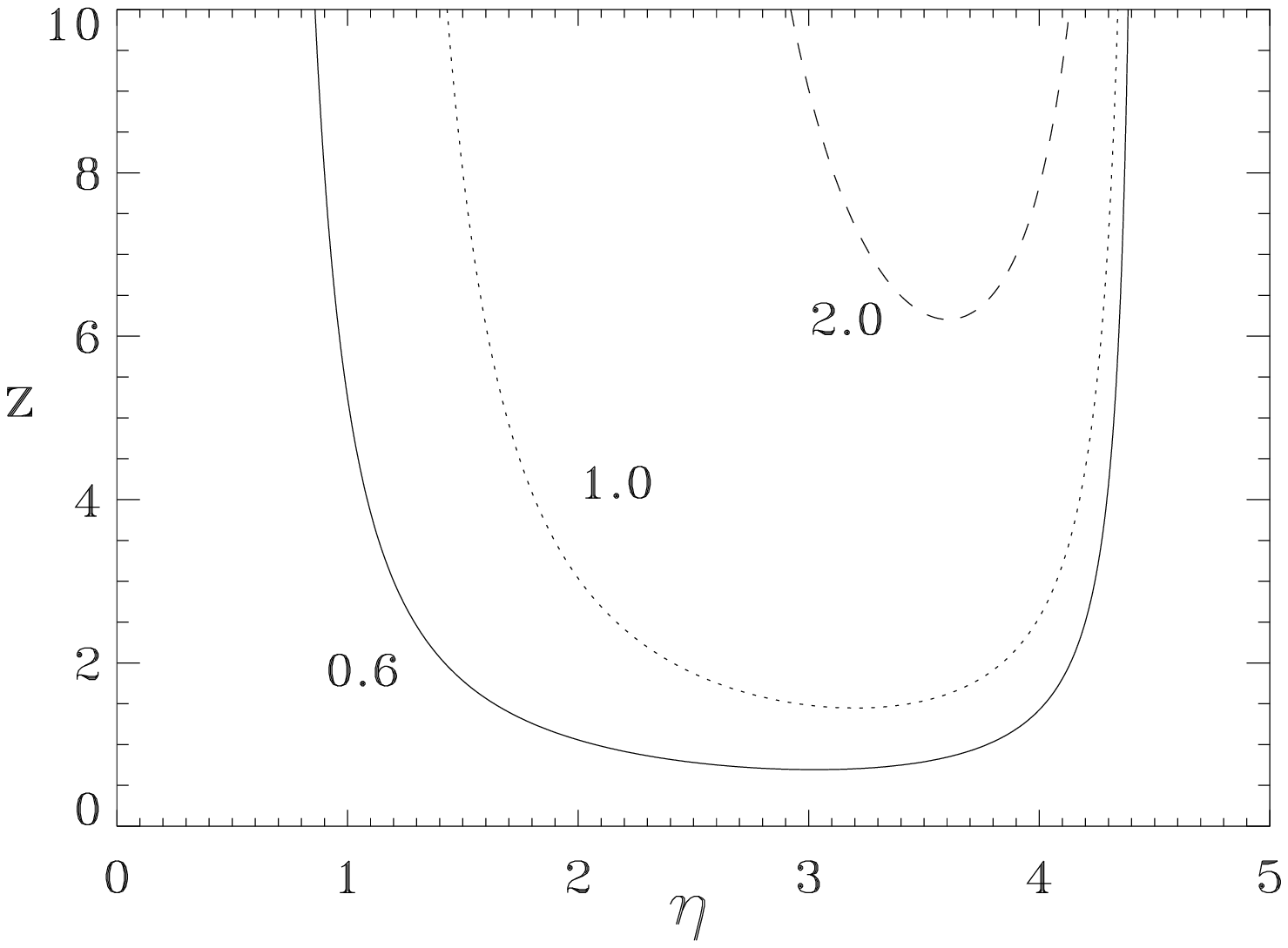}{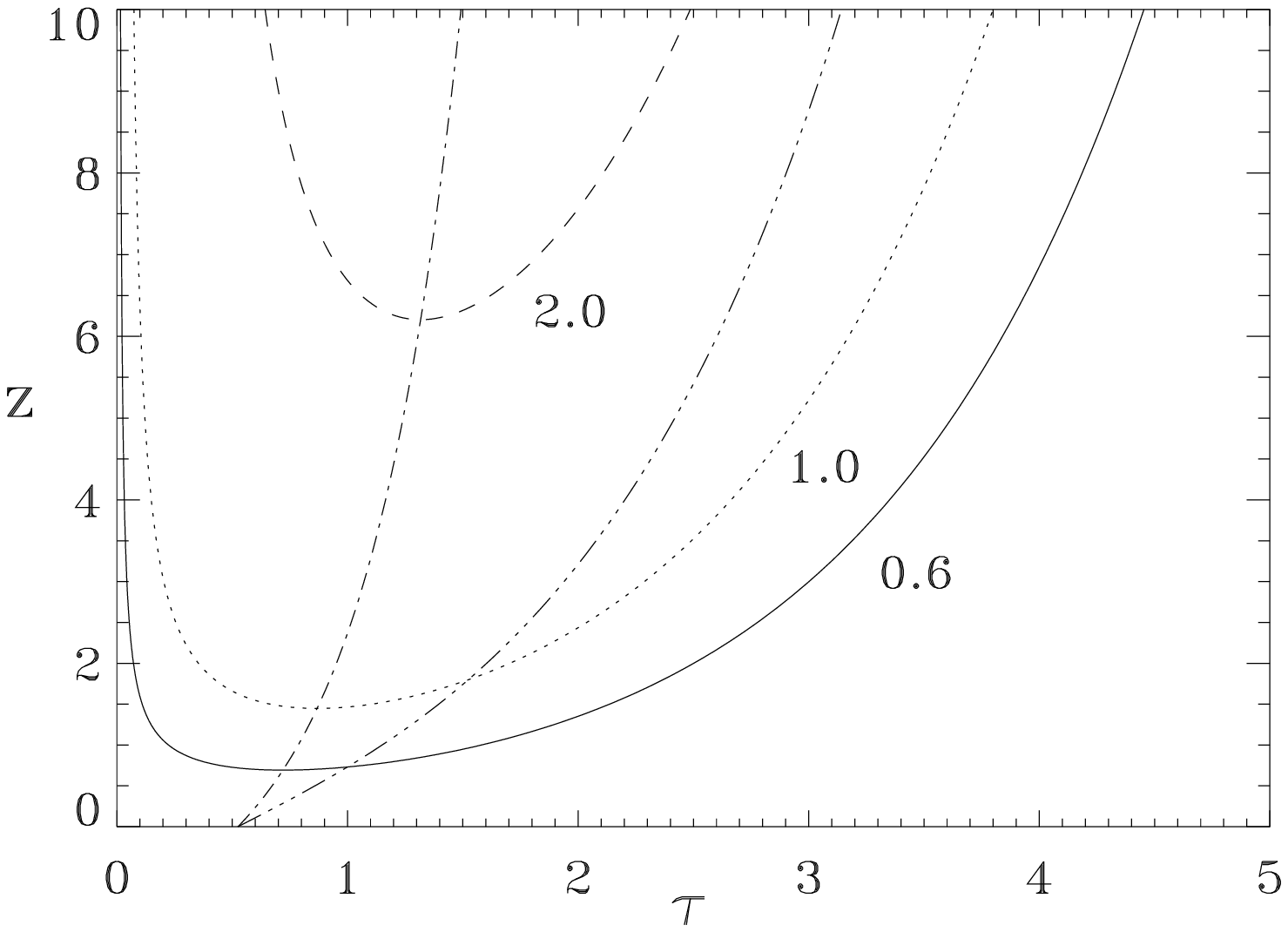}
\caption{Evolution of the redshift of three sources in the $\Lambda$-model. The sources are 
at $\chi = 0.60$ (solid curve), $1.0$ (dotted curve) and $2.0$ (dashed curve) and their 
worldlines are shown in Figure~\ref{fig3}b. 
(a) Redshift as a function of $\eta$, (b) Redshift as a function of $\tau$.
In b the redshifts $z_{\Lambda}$ (triple dot - dashed curve) and $z_{eq}$ (dot - dashed curve) 
are also shown as functions of $\tau$. See text for further explanations. \label{fig4}
}
\end{figure}
%%%%%%%%%%%%%%%%%%%%%%%%%%%%%%%%%%%%%%%%%%%%%%%

In Figure~\ref{fig4} we plot $z$ as a function of the time of observation for sources at various
distances, $\chi$. For simplification we have dropped 
the subscript $obs$ on both $\eta$ and $\tau$. 
We first ``see'' each source when it comes within our particle horizon, {\it i.e.}\ as
it was at the Big Bang ($\eta_{em} = 0$). It therefore enters our observable universe with
infinite redshift at cosmic time $\eta = \chi$. Before the cosmological constant becomes
dynamically important the redshift decreases more or less as in a universe without $\Lambda$
because the expansion is slowing down. 
As the effects of the cosmological constant begin to manifest themselves, the
source's redshift reaches a minimum,
and once $\Lambda$
completely dominates the expansion 
(i.e.\ for $\eta_{\bullet} < \eta \le \eta_{max}$ or
equivalently for $\tau_{\bullet} \ll \tau$; see
the discussion after eq.~[\ref{deh}]) 
the behaviour of the source's redshift with time is given by
\begin{equation}
\label{accz}
1 + z \approx 1 + \frac{\chi}{\eta_{max} - \eta} 
  \approx (1 + z_{\bullet})\;e^{\Omega_{\Lambda0}^{1/2}\;(\tau - \tau_{\bullet})}\;,
\end{equation}
where  $z_{\bullet} = z(\tau_{\bullet})$ and $\tau_{em} \leq \tau_\bullet$. 
For $\tau_\bullet < \tau_{em} < \tau$ we have that 
$1 + z = e^{\Omega_{\Lambda0}^{1/2}\;(\tau - \tau_{em})}$.
Hence all sources will redshift away
on a timescale $\Delta t \approx \Omega_{\Lambda0}^{- 1/2} t_{H0}$, the redshift going
to infinity at $\eta = \eta_{max}$ corresponding to $\tau = \infty$ 
when the scale factor becomes infinite.

In order to investigate this in more detail we introduce a redshift evolutionary timescale,
$T_z(\tau) =  [(d (1 + z)/ d\tau)/(1 + z)]^{-1}$, which can be compared to the
expansion timescale (normalized Hubble time), $T_a(\tau) = (H_0/H) = [(da/d\tau)/a]^{-1}$. 
Note that $T_z$ can take both positive and negative values, depending on whether the redshift is
increasing or decreasing.
By use of equation~(\ref{red}) we find for the FRW models in general that
\begin{equation}
\label{zder}
\frac{1}{T_z} = 
\frac{1}{(1 + z)} \frac{d (1 + z)}{d \tau} = 
   \frac{1}{a} \left[ \left(\frac{d a}{d \tau}\right)_{\tau} -  
   \left(\frac{d a}{d \tau} \right)_{\tau_{em}} \right]\;, 
\end{equation}
where the time derivative of $a$ is given by 
equation~(\ref{main}). Also note that in deriving this result we have used cosmic 
time dilation: $d \tau = (1 + z) d \tau_{em}$. In terms of the timescales, 
equation~(\ref{zder}) can be rewritten as
\begin{equation}
\label{timescales}
\frac{1}{T_z(\tau)} = \frac{1}{T_a(\tau)} -  
        \frac{1}{(1 + z) T_a(\tau_{em})} \;. 
\end{equation}

Next consider the  right hand side of equation~(\ref{timescales}) as a function of $z$.
It has a maximum at the redshift corresponding to a minimum value of $T_z$.
For Big Bang models with 
$\Omega_{\Lambda0} = \Omega_{Q0} = 0$ we have that $da/d\tau$ is a decreasing function 
of $\tau$, i.e.\ $d^2a/d \tau^2 < 0$, 
and thus $d(1 + z)/d\tau$ is negative during expansion. In such 
models the redshift of a given source always decreases with time (in recollapsing models 
the redshift eventually changes into blueshift, see e.g.\  \cite{bjo95}). 
However, for Big Bang models with a positive cosmological constant
the situation is different. Due to the dynamical effects of 
$\Lambda$, accelerated expansion starts at cosmic time $\tau_{\Lambda}$ given by
\begin{equation}
\label{second}
\left(\frac{d^2 a}{d \tau^2}\right)_{\tau_{\Lambda}} = 0
\end{equation}
and continues forever. It is clear from the discussion above 
that at a given cosmic time, $\tau > \tau_{\Lambda}$, 
the redshift corresponding to the epoch $\tau_{\Lambda}$ 
is also the redshift that minimizes $T_z$ and maximizes the change in redshift. 
This particular redshift, 
which we shall denote by $z_{\Lambda}$, thus locates the surface on the past light cone
which bounds the region where $\Lambda$ dominates
the expansion and within which the universe is accelerating. We shall refer to this
surface as the $\Lambda$-sphere.
Beyond the $\Lambda$-sphere the universe is still decelerating.

We next determine the redshift $z_{\Lambda}$. By use of equation~(\ref{main}) it is easy to show that
equation~(\ref{second}) is equivalent to the following algebraic equation for
$a$:
\begin{equation}
\label{algebra}
\frac{2 \Omega_{\Lambda0}}{ \Omega_{m0}}\; a^3 -
         \frac{(1 + 3w_Q) \Omega_{Q0}}{ \Omega_{m0}}\; a^{- 3 w_Q} = 1 \;,
\end{equation}
where we have assumed that $\Omega_{r0} = 0$. In the case $\Omega_{Q0} = 0$ 
the solution is $a = a_{\Lambda} = a(\tau_{\Lambda})$, where
\begin{equation}
\label{a2lambda}
a_{\Lambda} = \left(\frac{\Omega_{m0}}{2 \Omega_{\Lambda0}}\right)^{1/3}\;.
\end{equation}
At any time $\tau > \tau_{\Lambda}$ the observed redshift, $z_{\Lambda}$, 
of a source which emitted its light at $\tau_{\Lambda}$ is therefore given by
\begin{equation}
\label{zmin}
1 + z_{\Lambda} = \frac{a (\tau)}{a_{\Lambda}} = 
                  \left(\frac{2 \Omega_{\Lambda0}}{ \Omega_{m0}}\right)^{1/3}a(\tau) \;.
\end{equation}
In Figure~\ref{fig4}b the relation $z_{\Lambda} = z_{\Lambda}(\tau)$ is shown 
by the triple dot - dashed curve.
Note that equation~(\ref{zmin}) has a physical solution only if $a > a_{\Lambda}$.  
For our standard  $\Lambda$-model, we find that $a_{\Lambda} = 0.60$, 
corresponding to time $\tau_{\Lambda} = 0.52$, and $z_{\Lambda} = 0.67$.
Hence, in this model, cosmic acceleration started 
$\Delta t = (\tau_0 - \tau_{\Lambda}) t_{H0} = 4.3 h_0^{- 1}$ Gyrs ago. For $h_0 \approx 0.70$
this is $6.1$ Gyrs, and hence the acceleration started well before the formation 
of the solar system.

As time advances the $\Lambda$-sphere moves away from the observer and its 
conformal distance at time $\eta$ is given by
\begin{equation}
\chi = \eta - \eta_{\Lambda}\;,
\end{equation}
where $\eta_{\Lambda} = \eta (\tau_{\Lambda})$. Hence the  proper distance 
to the $\Lambda$-sphere is  
\begin{equation}
d_{\Lambda}(\eta) = R \chi = R_{H0} a(\eta)\; \frac{(\eta - \eta_{\Lambda})}{R_{H0}/ R_0}\;,
\end{equation}
and it is moving away from the observer with speed
\begin{equation}
v_{\Lambda} = \frac{d}{d t}(d_{\Lambda}) 
            = c \left(1 + \frac{(\eta - \eta_{\Lambda})}{R_{H0}/ R_0} 
                \left(\frac{d a}{d \tau} \right)  \right)\;.
\end{equation}
For our standard $\Lambda$-model we have that $\eta_{\Lambda} = 2.7$ and therefore 
$v_{\Lambda}(\tau_0) = 1.6 c$. Furthermore $d_{\Lambda}(\tau_0) = 0.56 R_{H0}$, which is about
47\% of the proper distance to the ultimate event horizon.
The evolution of the $\Lambda$-sphere is shown by the long dashed curve in Figure~\ref{fig2}. 
Note that this curve is the same as the observer's future light cone at 
time $\eta_{\Lambda}$.

From equations~(\ref{zder}) and (\ref{timescales}) 
we see that in a $\Lambda$ dominated universe $d (1 + z)/ d\tau =  dz/ d\tau$ 
is zero, and $T_a(\tau) = (1 + z)T_a(\tau_{em})$, 
when $(da/d\tau)_{\tau} = (da/d\tau)_{\tau_{em}}$. This corresponds to redshift
$z_{eq}$ given by
\begin{equation}
\label{zzero}
1 + z_{eq} = \left(\frac{\Omega_{\Lambda0}}{2 \Omega_{m0}}\right)\; a^3(\tau)
     \left( 1 + \sqrt{1 + \frac{4 \Omega_{m0}}{\Omega_{\Lambda0}\; a^3(\tau)}}\right) 
     =  \frac{1}{4}\left( \frac{a(\tau)}{a_{\Lambda}} \right)^3
     \left[ 1 + \sqrt{1 + 8 \left(\frac{a_{\Lambda}}{a(\tau)}\right)^3}\right]\;,
\end{equation}
where we have used equations~(\ref{zder}) and (\ref{main}). 
Note that $d (1 + z)/ d\tau$ 
(and hence $T_z$) is positive for $z < z_{eq}$ and negative for  $z > z_{eq}$.
The relation $z_{eq} = z_{eq}(\tau)$ is shown by the dot - dashed curve
in Figure~\ref{fig4}b for our standard $\Lambda$-model.
Note that equation~(\ref{zzero}) has a physical solution only if $a > a_{\Lambda}$.

From the discussion above we see that that although the $\Lambda$-sphere bounds the region 
in our visible universe where the cosmological constant dominates the expansion, 
the influence of $\Lambda$ extends beyond $z_{\Lambda}$ and is in principle observable 
approximately out to redshift $z_{eq}$. We shall return to this point in \S~\ref{changes}.

%------------------------------------------------------
\subsection{Brightness and Angular Size}
\label{brightsize}

At time $\tau$ the luminosity distance, $d_L$, of a source at conformal distance $\chi$ 
is given by
\begin{equation}
\label{dl}
d_L(\tau) = \left(\frac{L}{4 \pi F}\right)^{1/2} = R_0 r(\chi) (1 + z) a(\tau)  \;,
\end{equation}
where $L$ is the luminosity of the source, $F$ its apparent flux and $z$ its redshift. 
The relation $r = r(\chi)$ is given by equation~(\ref{rchi}). 

The distance modulus of the source is
\begin{equation}
\label{dm}
m - M = 25 + 5 \log{\left(\frac{d_L}{\rm Mpc}\right)} 
      = (5 \log{e}) \ln{\{r(\chi) (1 + z) a(\tau)\}} + {\rm constant}\;,
\end{equation}
where $\log$ stands for the logarithm with base $10$, 
$d_L$ is measured in Mpc, and $m$ and $M$ are the apparent and the absolute magnitude 
of the source respectively. 

The angular diameter distance of the source is given by
\begin{equation}
\label{dA}
d_A(\tau) = \frac{d_L(\tau)}{(1 + z)^2} = \frac{R_0 r(\chi) a(\tau)}{(1 + z)}
          =  R_0 r(\chi) a(\tau_{em}) \;,
\end{equation}
and if the source has proper diameter $D$, its apparent angular size $\phi$ is 
\begin{equation}
\label{phi}
\phi(\tau) = \frac{D}{d_A} 
                 = \left(\frac{D}{R_0}\right) \frac{1}{r(\chi)a(\tau_{em})}\;.
\end{equation}

In order to understand the evolution of $m$ and $\phi$ it is convenient to have 
an expression for their derivatives. By use of equations~(\ref{dm}), (\ref{phi}),
(\ref{zder}) and (\ref{timescales}) we find that
\begin{equation}
\label{dmag}
\frac{d m}{d\tau}  =  (5 \log{e})
\left(\frac{1}{(1 + z)}\frac{d (1 + z)}{d \tau} + \frac{1}{a}\frac{d a}{d \tau} \right)  = (5 \log{e})
   \left(\frac{1}{T_z(\tau)} +  \frac{1}{T_a(\tau)}\right)
\end{equation}
and
\begin{equation}
\label{dphi}
\frac{1}{\phi}\frac{d \phi}{d \tau} = 
\left(\frac{1}{(1 + z)} \frac{d z}{d \tau} 
   -  \frac{1}{a}\frac{d a}{d \tau}\right)  =
   \left(\frac{1}{T_z(\tau)} -  \frac{1}{T_a(\tau)}\right)
   = - \frac{1}{(1 + z) T_a(\tau_{em})}\;.
\end{equation}

%%%%%%%%%%%%%%%%%%%%%%%%%%%%%%%%%%%%%%%%%%
\begin{figure}[t]
\epsscale{1.}
\plottwo{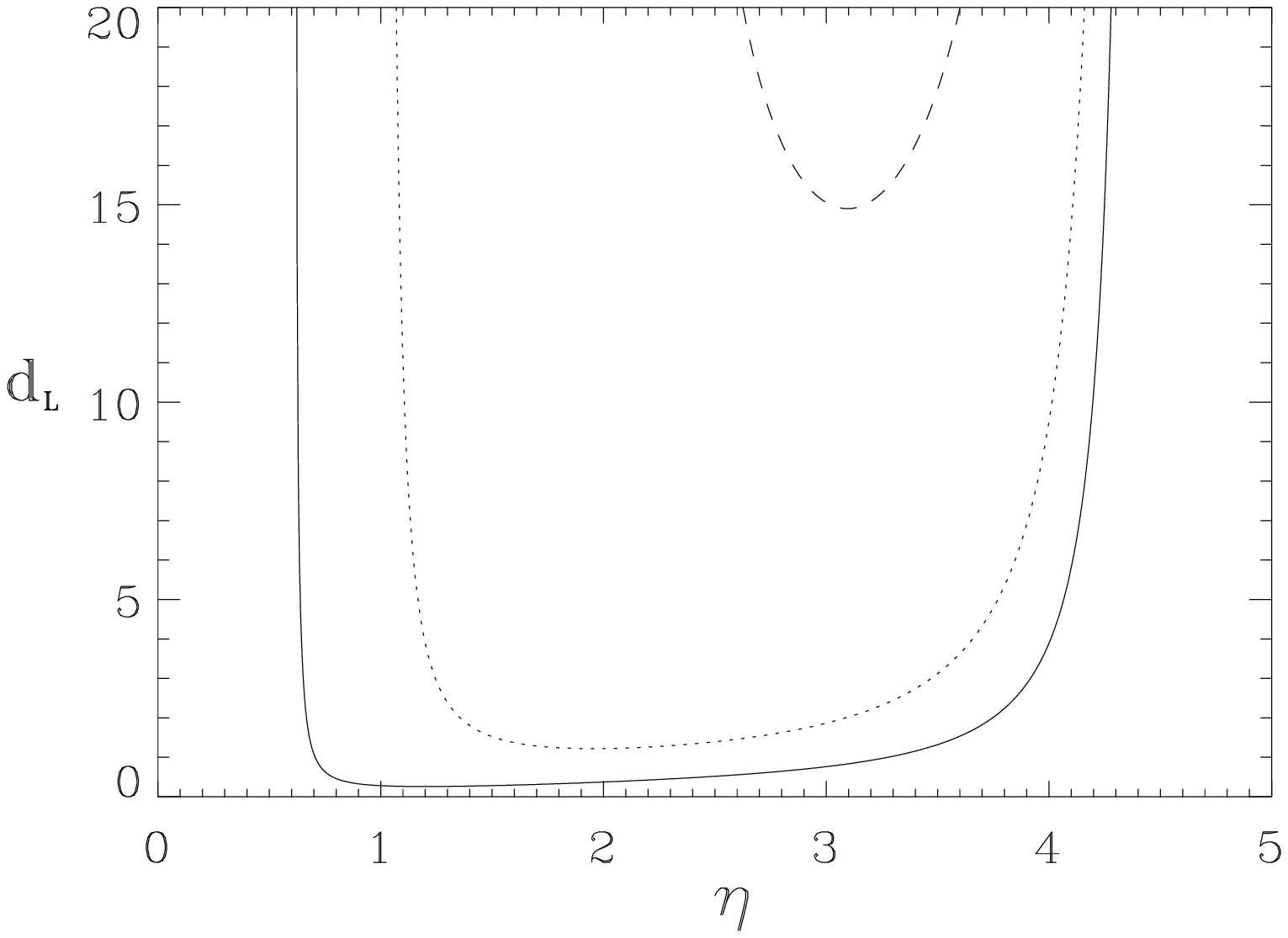}{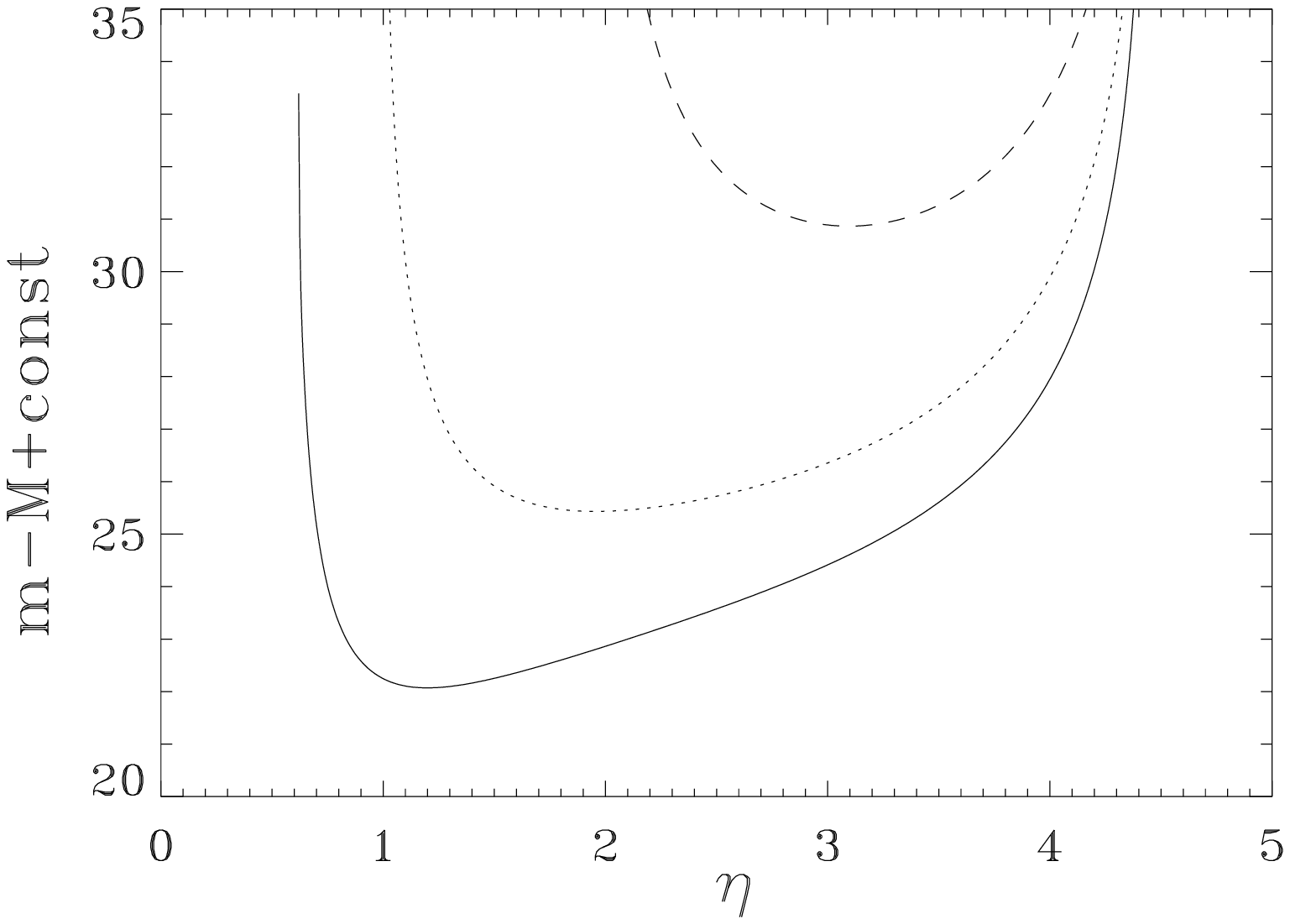}
\epsscale{1.}
\plotone{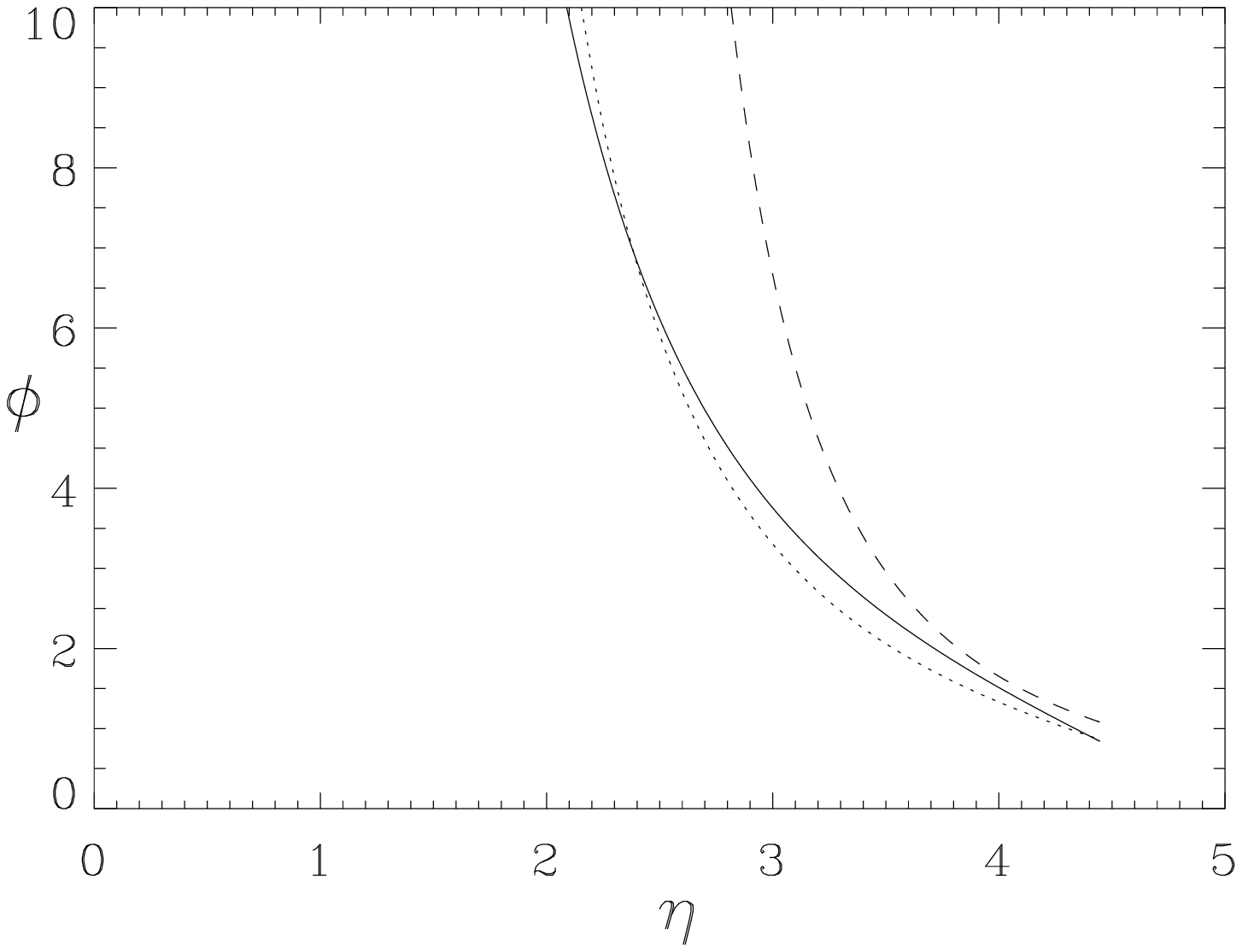}
\caption{(a) The luminosity distances (in units of $R_{H0}$) 
of the same sources as in Figure~\ref{fig4} 
as functions of ${\eta}$.
(b) The distance moduli of the sources as functions of ${\eta}$.
(c) The apparent angular sizes (in units of $D/R_{H0}$) as functions of ${\eta}$. 
\label{fig5}
}
\end{figure}
%%%%%%%%%%%%%%%%%%%%%%%%%%%%%%%%%%%%%%%%%%%%

In Figure~\ref{fig5}  
we show the luminosity distance, the distance modulus and 
the apparent angular size of sources at various distances $\chi$ as functions of cosmic time
in our standard $\Lambda$-model. 
The sources are assumed to have the same intrinsic luminosity
and the same proper diameters at all times ({\it i.e.}\ to be standard candles and standard rods).
A given source comes within our observable universe 
after a time determined by its comoving conformal distance, as long as it is less than
$\eta_{max}$, since sources further away are always beyond our event horizon and 
we never see them. 
Both $d_L$ and $m - M$  for the source
are infinite when the source appears 
with infinite redshift at $\eta = \chi$. 
These quantities decrease to a minimum and then increase to infinity at $\eta =
\eta_{max}$. Each source appears with infinite angular size at $\eta = \chi$, then decreases
monotonically with time to a minimum at $\eta = \eta_{max}$, the minimum
size being given by
\begin{equation}
\label{accphi}
\phi(\eta_{max}) = \left(\frac{D}{R_0}\right) \frac{1}{r(\chi)a(\eta_{max} - \chi)}
                 = \left(\frac{D}{R_{H0}}\right )\Omega_{\Lambda0}^{1/2} 
                   \frac{\chi}{r(\chi)}\;,
\end{equation}
corresponding to the final and finite angular diameter distance
\begin{equation}
d_A(\eta_{max}) = \frac{R_{H0}}{\Omega_{\Lambda0}^{1/2}}\; 
                   \frac{r(\chi)}{\chi}\;.
\end{equation}
In deriving these expressions we have used equation~(\ref{aetamax}).

Once the cosmological constant completely dominates the expansion 
(i.e.\ for  $\tau \gg \tau_{\bullet}$ corresponding to $\eta_{max} > \eta > \eta_{\bullet}$) 
the source's redshift increases according to equation~(\ref{accz}) and its luminosity distance
is given by
\begin{equation}
\label{bdl}
d_L(\tau) \sim  e^{2 \Omega_{\Lambda0}^{1/2}\tau}\;.
\end{equation}
Hence its apparent magnitude grows linearly with time:
\begin{equation}
\label{bm}
m(\tau) \approx  {\rm constant} + 10 (\log{e})\; \Omega_{\Lambda0}^{1/2}\;\tau\;.
\end{equation}
It is therefore clear that once
a source has entered our observable universe we never see it leave, 
although once cosmic acceleration has started, its brigtness gets 
below any finite detection limit in a few Hubble times.

%------------------------------------- 5 --------------------
\section{CHANGES IN OBSERVABLE PROPERTIES AT THE PRESENT EPOCH}
\label{changes}

The main astronomical evidence for cosmic acceleration comes from investigations based on
classical cosmological tests, in particular the $m - z$ relation
with Type Ia supernovae as standard candles. This has been treated in
great detail by the research groups who discovered the acceleration  
\citep{per99,gar98,rie98,arie01} and we shall therefore not discuss these tests here.

A related but more difficult approach, at least observationally, 
is to consider changes in cosmological observables over
extended periods of observing time and see how they are affected by a cosmological constant
or a quintessence field. 
We have laid the foundation for such a discussion in the previous section.

Let us consider a source with redshift $z$ on our present past light cone and determine its
change in redshift, $\Delta z$, during a time interval $\Delta \tau_0 \ll \tau_0$. 
Using equation~(\ref{zder})  we find
\begin{equation}
\frac{1}{(1 + z)} \frac{\Delta z}{\Delta \tau_0}  = 
         \frac{1}{(1 + z)}\; \frac{\Delta (1 + z)}{\Delta \tau_0}  \approx  
         \left(1 - \left(\frac{d a}{d \tau}\right)_{\tau_{em}}\right)\;,
\end{equation}
and by equation~(\ref{main}) this can be written in terms of observables as
\begin{equation}
\label{deltaz}
\frac{1}{(1 + z)} \frac{\Delta (1 + z)}{\Delta \tau_0} \approx  \left( 
1 -  \left\{\Omega_{m0}(1 + z) + \Omega_{r0}(1 + z)^2  
           + \frac{\Omega_{\Lambda 0}}{(1 + z)^2} 
           + \frac{\Omega_{Q0}}{(1 + z)^{-(1 + 3w_Q)}} 
           + (1 - \Omega_0)\right\}^{1/2}\right)\;,
\end{equation}
where $\Omega_0 = \Omega_{m0} + \Omega_{r0} + \Omega_{\Lambda0} +  \Omega_{Q0}$.

%%%%%%%%%%%%%%%%%%%%%%%%%%%%%%%%%%%%%%
\begin{figure}[t]
\epsscale{1.}
\plottwo{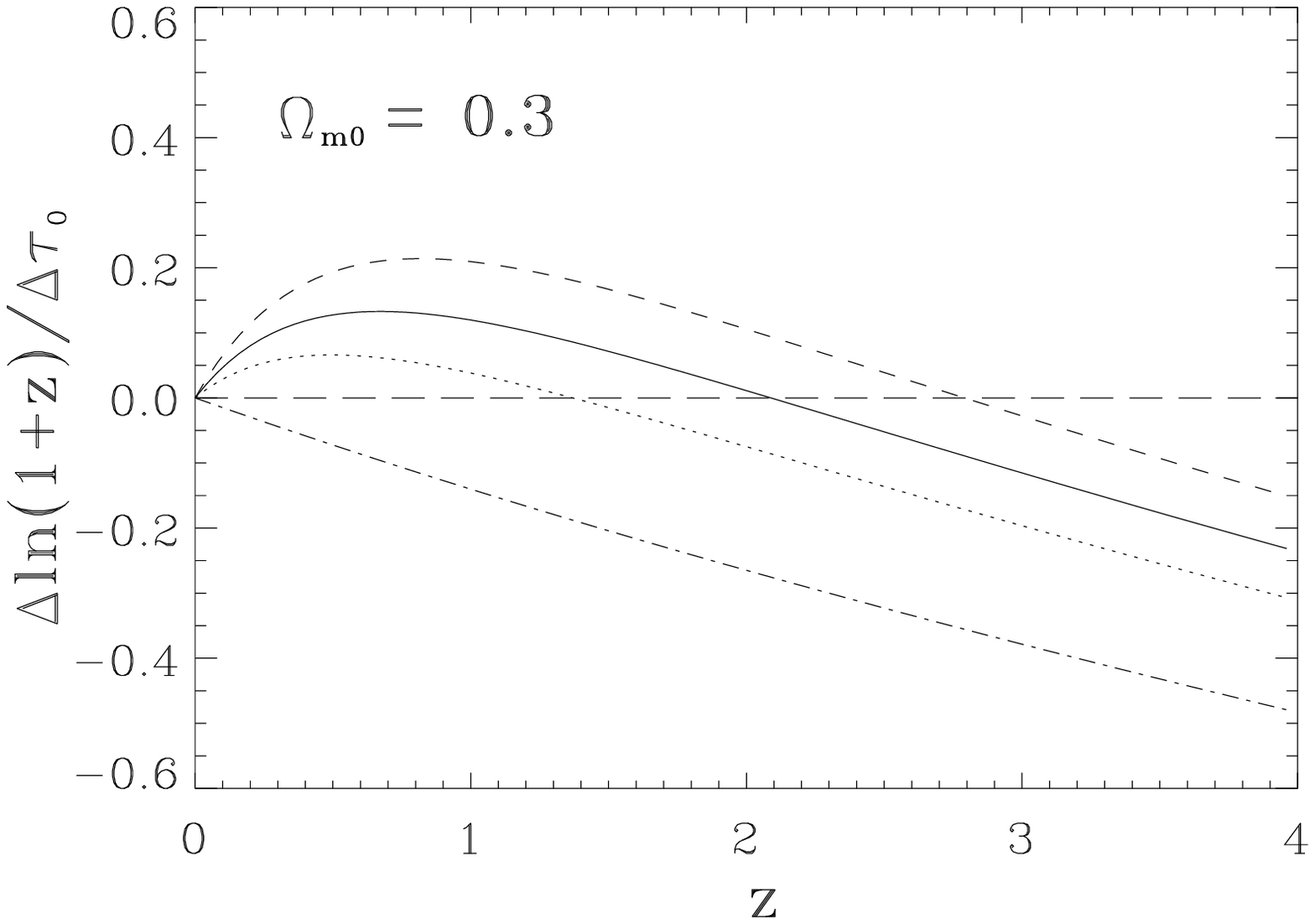}{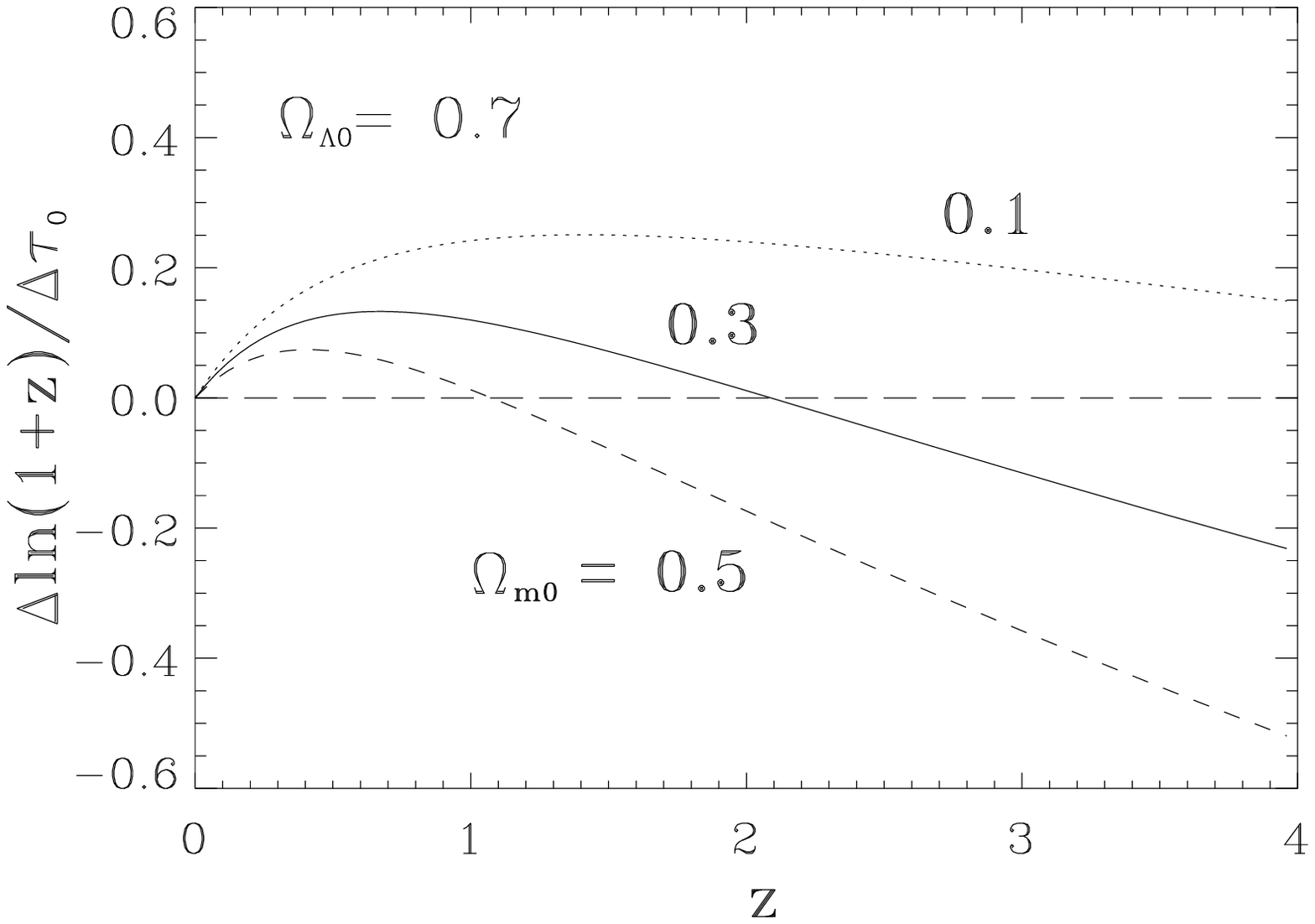}
\caption{(a) $(\Delta (1 + z)/ \Delta \tau_0)/(1 + z)$
as a function of $z$ for $\Omega_{m0} = 0.30$ and $\Omega_{\Lambda0}= 0$ (dash - dotted curve) 
$\Omega_{\Lambda0}= 0.50$  (dotted curve), $0.70$ (solid curve) and $0.90$ (dashed curve).
(b) The same quantity as a function of $z$
for $\Omega_{\Lambda0}= 0.70$
and  $\Omega_{m0} = 0.10$ (dotted curve), $0.30$ (solid curve) and $0.50$ (dashed curve).
All models in this figure have  $\Omega_{Q0} = \Omega_{r0} = 0$.
\label{fig6}
}
\end{figure}
%%%%%%%%%%%%%%%%%%%%%%%%%%%%%%%%%%%%%%%

In Figure~\ref{fig6} we plot $(\Delta (1 + z)/ \Delta \tau_0)/(1 + z)$ as a function of $z$
for selected values of $\Omega_{m0}$ and $\Omega_{\Lambda0}$ with 
$\Omega_{Q0} = \Omega_{r0} = 0$. For this choice of parameters
the zeros are at $z = 0$ and $z = z_{eq}$, where by equation~(\ref{zzero})
\begin{equation}
1 + z_{eq} = \left(\frac{\Omega_{\Lambda0}}{2 \Omega_{m0}}\right)
\left( 1 + \sqrt{1 + \frac{4 \Omega_{m0}}{\Omega_{\Lambda0}}}\right) \;.
\end{equation}
The presence of the cosmological constant makes $(\Delta (1 + z)/ \Delta \tau_0)/(1 + z)$ 
positive for $0 < z < z_{eq}$ and its maximum value at the $\Lambda$-sphere
(with $1 + z_{\Lambda} = (2\Omega_{\Lambda0}/ \Omega_{m0})^{1/3}$) is given by
\begin{equation}
\label{change}
\left(\frac{1}{(1 + z)}\; \frac{\Delta (1 + z)}{\Delta \tau_0}\right)_{z = z_{\Lambda}}
   = 1 - \left\{3 \left[ \left(\frac{\Omega_{m0}}{2}\right)^2 \Omega_{\Lambda0} \right]^{1/3} 
     + (1 - \Omega_{m0} - \Omega_{\Lambda_0}) \right\}^{1/2} \;.
\end{equation}

For our standard  $\Lambda$-model 
we find that $z_{eq} = 2.1$ and that the maximum change given 
by equation~(\ref{change}) is $0.13$ (corresponding to $T_z = 7.7$) at $z_{\Lambda} = 0.67$. 
Hence the maximum effects of the cosmological constant 
on the change in redshift is given by 
$\Delta z = 1.67 \times 0.13 \Delta \tau_0 = 0.22 \Delta t_0 /t_{H_0}$. 
For $\Delta t_0 = 100$ years, say, 
$\Delta z$ at maximum is therefore only of the order of $10^{-9}$. 
This is a very small number and such
minute changes in $z$ will probably not be observable in the near future 
(see however \cite{loe98} for a detailed discussion 
of the observational situation).

In a similar way one can investigate changes in the apparent magnitude and the 
apparent angular size of a given source with present redshift $z$. 
Using equations~(\ref{dmag}) and (\ref{dphi}) 
we find to first order in $\Delta \tau_0$:
\begin{equation}
\label{deltam}
\frac{1}{(5 \log{e})}\frac{\Delta m}{\Delta \tau_0}  \approx 
\left(1  + \frac{1}{(1 + z)}\frac{\Delta (1 + z)}{\Delta \tau_0}\right) 
\end{equation}
and
\begin{equation}
\label{deltaphi}
\frac{1}{\phi}\frac{\Delta \phi}{\Delta \tau_0}  \approx 
\left(\frac{1}{(1 + z)} \frac{\Delta (1 + z)}{\Delta \tau_0} - 1\right)  \;,
\end{equation}
with $(1 + z)^{-1}({\Delta (1 + z)}/{\Delta \tau_0})$ given by equation~(\ref{deltaz}).
It is clear that Figure~\ref{fig6} can be used to investgate the behaviour of both 
$\Delta m / \Delta \tau_0$ and $\Delta \phi / \phi \Delta \tau_0$ 
as functions of $z$ for the selected values of $\Omega_{m0}$ and $\Omega_{\Lambda 0}$. 
From the results already obtained we find for our standard $\Lambda$-model 
that at $z_{\Lambda}$ the change in magnitude
is $\Delta m = 2.5 \Delta \tau_0 \approx 10^{-8}$ over a period of 100 years. 
In a similar way the maximum relative
changes in $\phi$ due to the accelerated expansion is 
$\Delta \phi / \phi = - 0.87 \Delta \tau_0 \approx - 10^{-8}$ over the same perod
of time. Such  small changes will presumably not be observable in the near future. 

We conclude this section by emphasizing that
although the prospects for actually measuring the changes in redshift, apparent brightness 
and apparent size of
cosmological sources do not seem promising, it is of interest to investigate 
Figure~\ref{fig6} with respect to the general effects of a cosmological constant at the present
epoch. We see e.g.\ that for distant sources with high $z$ ({\it i.e.}\ $z > z_{eq} > z_{\Lambda}$), 
corresponding to emission at cosmic times before $\Lambda$
becomes dynamically dominant, their redshift is decreasing and their apparent brightness
is also decreasing, but relatively slowly. However, for cosmological sources with $z < z_{eq}$ 
the redshift is increasing due to the repulsive effects of $\Lambda$, 
and the apparent brightness is decreasing relatively fast. Thus the influence of $\Lambda$ is
considerable outside the $\Lambda$-sphere, out to $z \approx z_{eq}$.
This is in agreement with the time evolution of the  observables 
discussed in \S~\ref{redbrightsize} and \ref{brightsize} and demonstrated 
in Figures~\ref{fig4} and~\ref{fig5}.

%-------------------------------- 6 -------------------------------
\section{CAUSAL CONNECTIONS AND THE EXTENT OF OUR OBSERVABLE UNIVERSE}
\label{causal}

\begin{figure}[t]
\epsscale{1.}
\plottwo{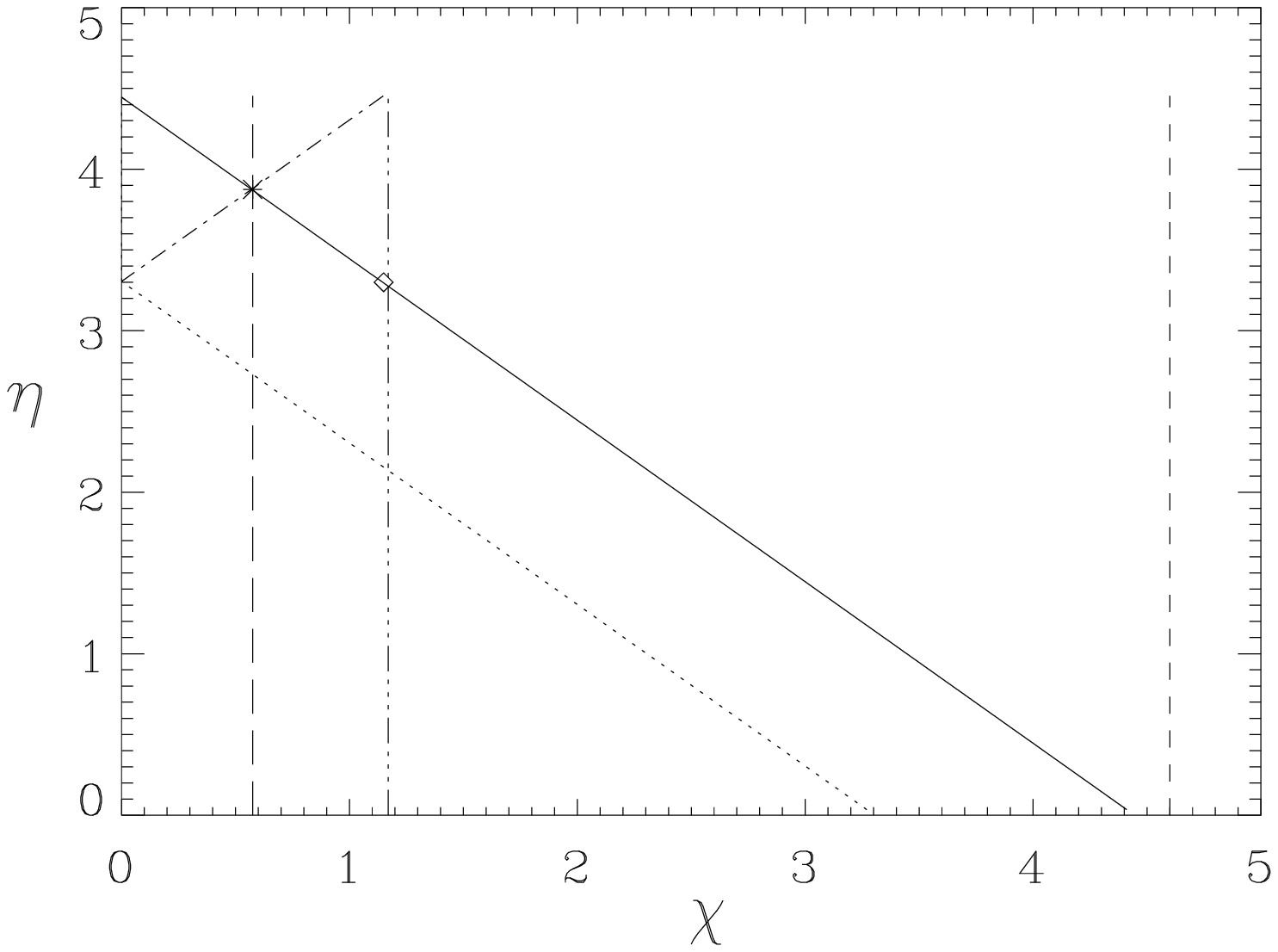}{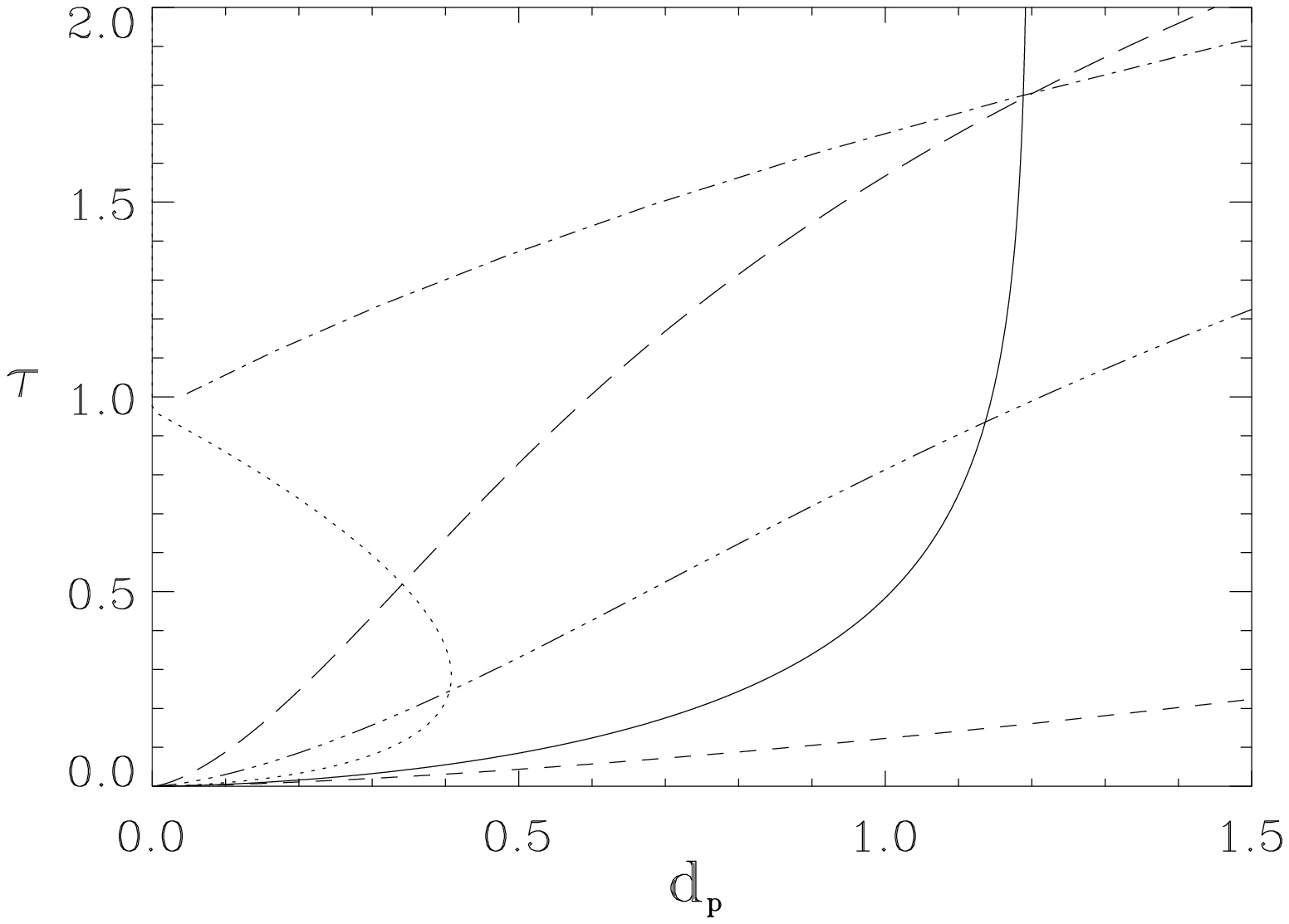}
\caption{(a) Sources at $\chi_{\star}$ (long dashed line) and 
$\chi_c$ (dash-triple dotted line) 
crossing the event horizon at times ${\eta}_{\star}$ and ${\eta}$ respectively.
Also shown is the world line of a source which is always outside the event horizon 
(short dashed line). (b) The same situation in proper coordinates $\tau$ vs.\ $d_p = d/R_{H0}$.
The model is our standard $\Lambda$-model with the time of observation equal to 
$\eta_0 = 3.3$. See text for further explanations. \label{fig7}
}
\end{figure}

Consider Figure~\ref{fig7}. The future light cone (dot-dashed curve) 
at observing time $\eta$ crosses the 
event horizon (solid curve) at the event ($\chi_{\star}$, $\eta_{\star}$). 
The world line of a source
that passes through that event is also shown (long dashed curve). 
We will not be able to receive 
any signal from that source sent later than $\eta_{\star}$. Because of symmetry, 
the source can not receive any signal from us sent after $\eta_{\star}$.
Hence the source passes out of our 
sphere of influence at time $\eta_{\star}$. However, our evolving past light cone 
(shown as it is at the present epoch by the dotted curve) crosses
the source's world line right till the end  at $\eta_{max}$, at which 
time the source is seen  as it was at $\eta_{\star}$ but with infinite redshift.
At time $\eta$ we  see the source with redshift $z_{\star}$ 
as it was at time $\eta_{em \star}$. It is easy to see that
\begin{equation}
\eta_{\star} = \frac{\eta_{max} + \eta}{2}
\end{equation}
and
\begin{equation}
\chi_{\star} = \frac{\eta_{max} - \eta}{2}\;.
\end{equation}
Also
\begin{equation}
\eta_{em \star} = \frac{3 \eta - \eta_{max}}{2}\;,
\end{equation}
and for this last equation to be valid we must have $\eta \geq \eta_{max}/3$, since
at earlier times the source is outside the  particle horizon. It's redshift
is given by
\begin{equation}
1 + z_{\star} = \frac{a(\eta)}{a(\eta_{em \star})}\;.
\end{equation}

%%%%%%%%%%%%%%%%%%%%%%%%%%%%%%%%%%%%%%%%%%%%%%%%
%%%%%%%%%%%%%%%%%%%%%%%%%%%%%%%%%%%%%%%%%%%%%%%%

Next consider the source at $\chi_c$ in Figure~\ref{fig7} (dash-triple dotted curve) 
that is crossing the event horizon at time $\eta$. Clearly
\begin{equation}
\chi_c = \eta_{max} - \eta = 2 \chi_{\star}\;,
\end{equation}
and we see this source with redshift $z_c$ given by
\begin{equation}
1 + z_c =  \frac{a(\eta)}{a(\eta_{em c})} \;.
\end{equation} 
the light we see being emitted at time 
\begin{equation}
\label{em0}
\eta_{em c} = 2 \eta - \eta_{max}\;.
\end{equation}
Note that we must have $\eta > \eta_{max}/2$ for equation~(\ref{em0}) to be valid. 
At earlier times the source is outside the particle horizon.

For our standard world model shown in Figure~\ref{fig7} we have that the present time of 
observation, $\eta$, is equal to ${\eta}_0 = 3.3$. Also  
${\eta}_{max} = 4.5$, 
${\eta}_{\star} = 3.9$, $\chi_{\star} = 0.57$ and ${\eta}_{em\star} = 2.7$ 
(we remind the reader that numerical values of all $\eta$'s and  
$\chi$'s are in units of $R_{H0}/R_0$). 
We also find that $z_{\star} = 0.68$. This means, that the light being emitted now by
sources having present redshift greater than $0.68$, will not reach us 
until the sources have crossed our event horizon, 
at which time they are completely out of causal contact.
 
In this model we also have $\chi_c = 1.1$, ${\eta}_{emc} = 2.2$ and 
$z_c = 1.7$. Hence all sources with present redshift greater than $1.7$ have already crossed
our event horizon, and are thus completely out of causal contact with us
(see also \cite{sta99} and \cite{stab99} who reach similar conclusions).
From this one can easily estimate the number of sources that are still within the event horizon  
as compared with the initial matter content. The number of sources is proportional to the
comoving 
proper volume and hence the relative number of sources presently within the horizon is equal to
$(\chi_c(\eta_0) / \chi_{eh}(0))^3 = (\chi_c(\eta_0) / \eta_{max})^3 = 
0.015$. This means that more than
98\% of all sources initially within our 
observable part of the universe have already crossed the event horizon.

We can also estimate the portion of observable sources that we could
have seen by now, at least in principle.
It is simply given by $(\chi_{ph}(\eta_0) / \chi_{eh}(0))^3 = (\eta_0 / \eta_{max})^3
\approx 0.40$. Hence,
we still have not seen 60\% of the observable sources.

To summarize: In our standard $\Lambda$-model the event horizon will ultimatly be stationary 
at a proper distance of $1.2 R_{H0} = 3.6 h_0^{-1}$ Gpc (note that this 
is also the ultimate angular diameter distance to the event horizon). 
We have already seen about 40\% of the sources that are in principle observable, 
but about 98\% of all cosmic sources originally within our observable 
part of the universe have already left.
This includes all sources that we presently see and have redshift higher than $1.7$. 
Because of the finite speed of light we will eventually be able to see all
the sources originally inside our event horizon, 
but only as they were in the past before crossing the horizon.
Once the cosmological constant 
dominates the expansion their redshift will, however, increase
exponentially with time and they will fade away on a timescale measured in a few Hubble times.
Of course this applies only to sources that are distant enough to participate in the cosmic 
expansion, {\it i.e.}\ sources outside our local supercluster but originally 
within our event horizon.

%---------------------------- 7 ----------------------------------------
\section{QUINTESSENCE}
\label{quintessence}

Although by invoking Einstein's cosmological constant it is easy to explain why 
the expansion of the universe is presently accelerating,
there has recently been considerable discussion of the possibility
that something else is causing the acceleration, mimicking the effects of a
cosmological constant at the present epoch. The main reason for this is simply
that theoretical particle physicists have so far been unable
to explain why the cosmological constant has such a small nonzero value 
($\Lambda \approx 10^{- 56}~{\rm cm}^{-2}$ corresponding to 
$\Omega_{\Lambda0} \approx 0.7$), their most 
natural estimates giving values of $\Lambda \sim 10^{64} ~{\rm cm}^{-2}$ corresponding to
$\Omega_{\Lambda0} \sim 10^{120}$
(see e.g.\ \cite{wei89,wei00} and \cite{wit00} and references therein). 

Several other possible causes for the cosmic acceleration have been discussed in the literature, 
including quintessence, frustrated network of topological defects, 
time-varying particle masses and effects 
from extra dimensions (see e.g.\ \cite{hut00}, \cite{bin00} and \cite{wei01} 
for further discussion and references).

In this paper we shall only investigate the effects of one of these alternatives. We choose
the case of a slowly evolving scalar field (or quintessence; 
\cite{pee88,rat88,wet88,cal98,zla99}) leading to an equation of state of
the form $P_Q = w_Q \rho_Q c^2$ where $w_Q$ may or may not be a function of cosmic time. 
We assume as before that we are investigating epochs where radiation can be 
neglected ($\Omega_{r0} = 0$) and that the quintessence
field is decoupled from matter. 

In order to investigate the effects of quintessence on the evolution of 
the observable universe
we shall furthermore assume that $\Omega_{\Lambda0} = 0$ and that $w_Q$ is a constant
(assuming a constant $w_Q$ is not a serious restriction since our results can easily be 
extended to the time dependent case). It then follows from equation~(\ref{eq2})
that the mass-energy density of the quintessence component is given by 
\begin{equation}
\rho_Q = \rho_{Q0}\;a^{- 3(1 + w_Q)}\;.
\end{equation}
Realistic expanding models, with $\rho_Q$ decreasing with time, thus require $w_Q > -1$. 
The case corresponding to  $w_Q = -1$ is of course equivalent to the 
cosmological constant case.

\begin{figure}[t]
\epsscale{1.}
\plottwo{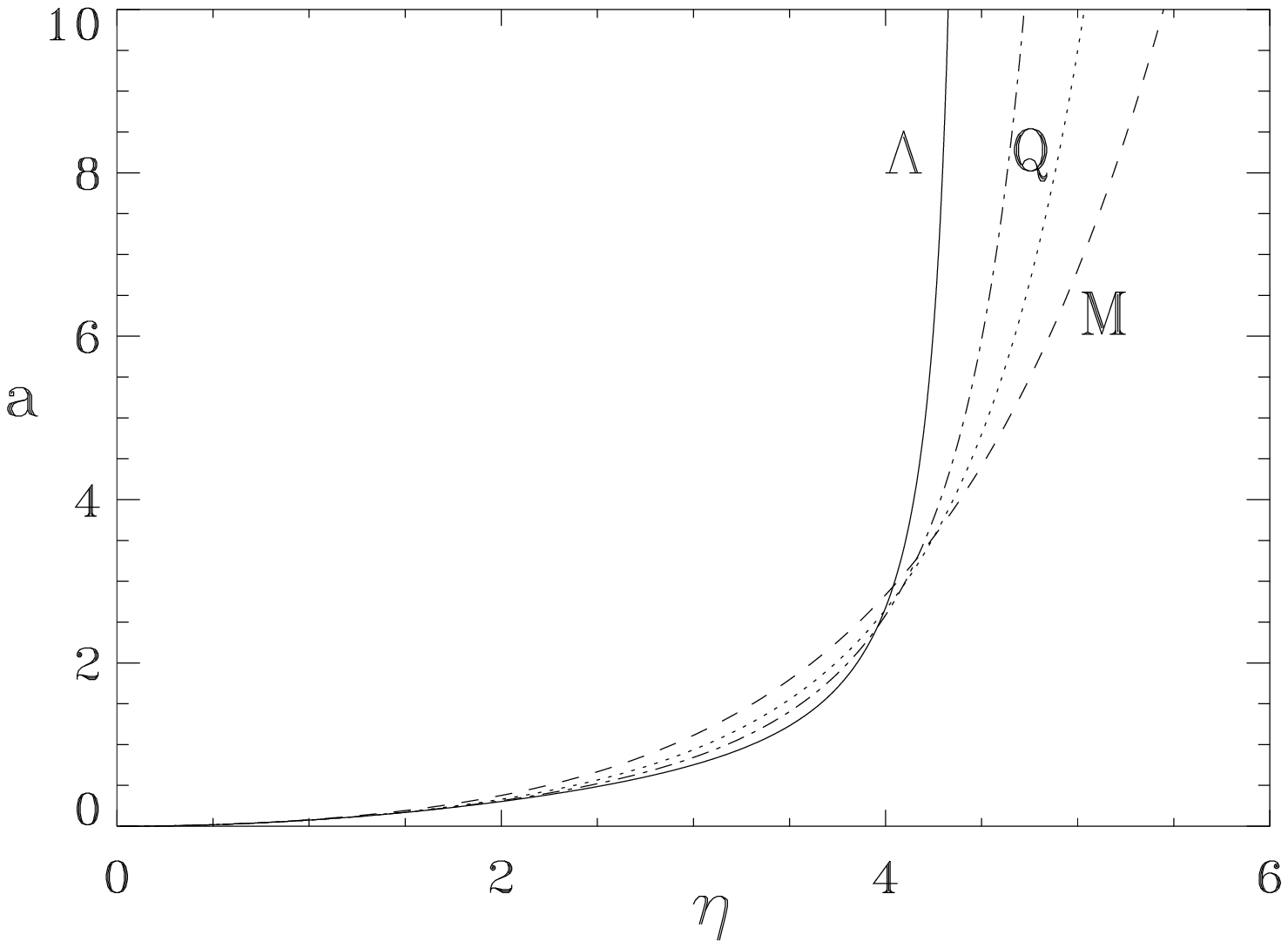}{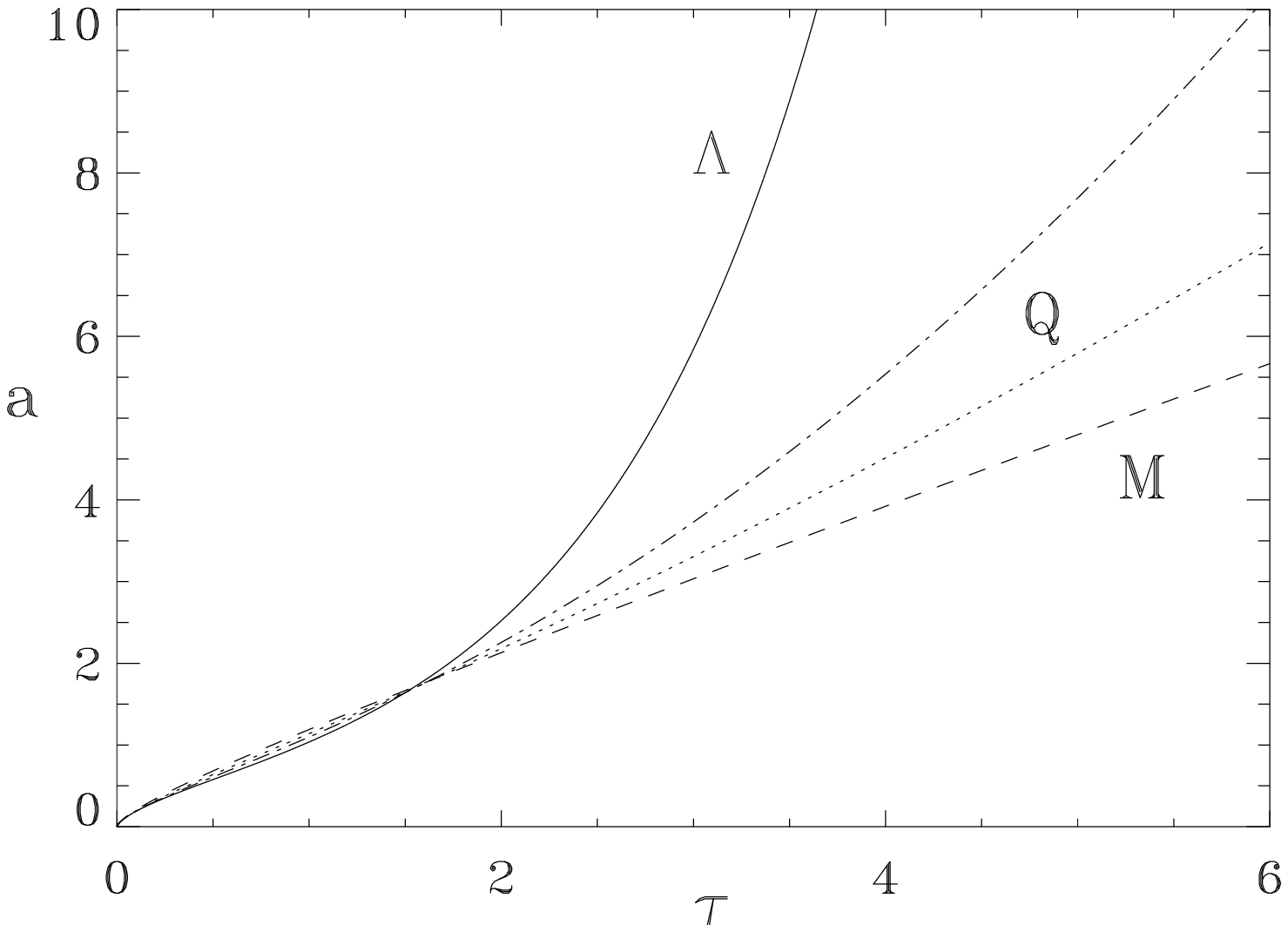}
\caption{Comparison of the time evolution of the scale factor, $a$, in
different cosmological models. (a) $a$ as a function of $\eta$  
for the standard $\Lambda$-model 
(solid curve, marked $\Lambda$) and two quintessence models with  
$\Omega_{m0} = 0.30$ and $\Omega_{Q0} = 0.70$. One has  $w_Q = - 0.50$ 
(dotted curve, marked $Q$), the other $w_Q = - 2/3$ (dash - dotted curve). 
Also shown is the $M$-model with 
$\Omega_{m0} = 0.30$ and  $\Omega_{\Lambda0} =  \Omega_{Q0} = 0$ 
(dashed curve, marked $M$).
(b) $a$ as a function of $\tau$ for the same models. 
Compare with Fig.~\ref{fig1}. \label{fig8}
}
\end{figure}

Equation~(\ref{main}) now reduces to
\begin{equation}
\label{mainq}
\frac{d a}{d \tau} = \left\{\Omega_{m0} \left(\frac{1}{a} - 1 \right)  
     + \Omega_{Q0}\left(\frac{1}{a^{1 + 3 w_Q}} - 1 \right) + 1\right\}^{1/2}\;,
\end{equation}
from which we see that for
ever-expanding Big Bang models, the quintessence field will ultimately
dominate the expansion if $w_Q < 0$, 
and at late times the scale factor will grow with 
$\tau$ according to 
\begin{equation}
a \sim (n\; \Omega_{Q0}^{1/2}\; \tau)^{n}\;, 
\end{equation}
where
\begin{equation}
n = \frac{2}{3(1 + w_Q)}\;.
\end{equation}
This should be compared with the corresponding behaviour of $a$ at late times in a 
$\Lambda$-dominated universe (eq.~[\ref{alambda}]). Examples demonstrating
the difference are shown in Figure~\ref{fig8}. 

From the results above and the discussion at the end 
of \S~\ref{eventhor} we see that only models with $w_Q < - 1/3$ (corresponding to $n > 1$) 
have event horizons. The theoretically interesting range for $w_Q$ is 
therefore $- 1 < w_Q < - 1/3$ (corresponding to $\infty > n > 1$). 
This should be compared to observations, which seem to indicate that 
quintessence can only be viable at present if $- 1 < w_Q < w_{max}$ with 
the upper limit, $w_{max}$, not greater than $- 0.4$ and probably lower
(see e.g.\ \cite{hut00} and \cite{wan00} for a discussion of 
the observational situation). Thus, if the cosmic acceleration is due to
quintessential dark energy, the universe has event horizons 
just as in the case of a true cosmological constant (see in this context \cite{hel01} 
and  \cite{fis01}, who discuss the problems this poses for string theory). 
It should be pointed out, however, that if for
some reason $w_Q$ changes in the future, so that it ultimately becomes larger than $- 1/3$
(corresponding to $n < 1$) then the event horizons disappear.
Also note that  for a given $\Omega_{Q0}$ the maximum conformal time,
$\eta_{max}$, tends to infinity as $w_Q$ approaches $ -1/3$ from below.

In what follows we shall investigate in detail the evolution of the observable universe 
when quintessence is the single cause of cosmic acceleration. 
We use the same methods as above and compare the results to those of models with 
a positive cosmological constant.

%%%%%%%%%%%%%%%%%%%%%%%%%%%%%%%%%%%%%%%%%%
%%%%%%%%%%%%%%%%%%%%%%%%%%%%%%%%%%%%%%%%%%%

In Figure~\ref{fig8} we show the normalized scale factor $a$ as a function of 
time for $\Omega_{m0} = 0.30$, $\Omega_{Q0} = 0.70$ and two values of $w_Q$: 
$ - 0.50$ (dotted curve, marked $Q$) and $- 2/3$ (dash - dotted curve).
In the following we shall refer to  the model with  $w_Q= - 0.50$ as the $Q$-model. 
For this model
$\eta_{max} = 7.7$ and $\eta_0 = 3.1$. 
Also shown for comparison are two other models: our standard $\Lambda$-model 
(solid curve, marked $\Lambda$), 
and a model, which we shall call the $M$-model, 
with $\Omega_{m0} = 0.30$ and both $\Omega_Q$ and $\Omega_{\Lambda0}$ equal 
to zero (dashed curve, marked $M$). 
Note that since $w_Q < - 1/3$ for the quintessence models, they both have event horizons.

From Figure~\ref{fig8} one sees  how the expansion of quintessence models, 
with $\Omega_{Q0}$ and $\Omega_{m0}$
both fixed, depends on $w_Q$. As $w_Q$ takes decreasing values in the range 
from $- 1/3$ to $- 1$, the 
corresponding curves  change continuously from the $M$-curve to the $\Lambda$-curve.

%%%%%%%%%%%%%%%%%%%%%%%%%%%%%%%%%%%%%%%%%%
\begin{figure}[t]
\epsscale{1.}
\plottwo{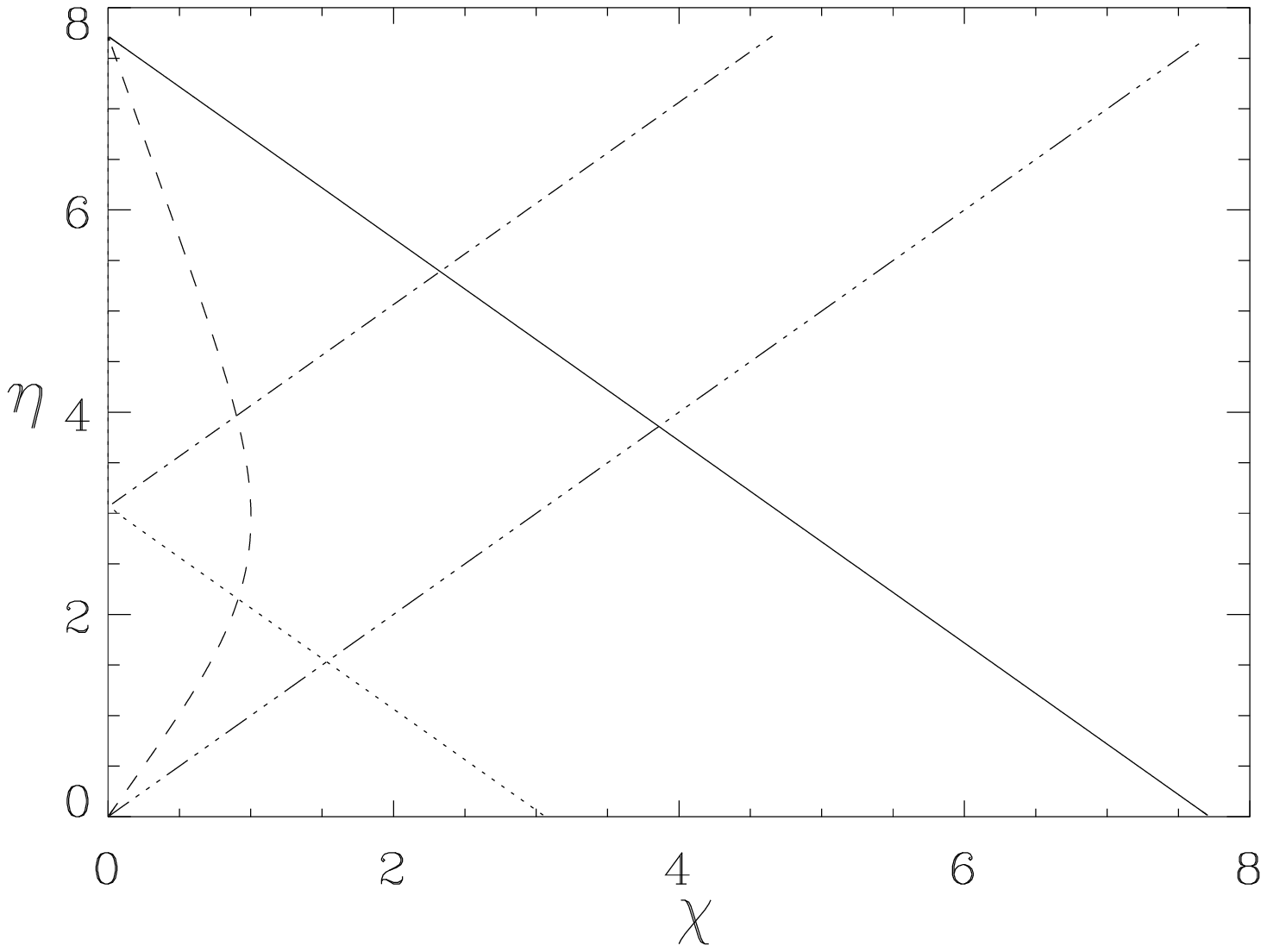}{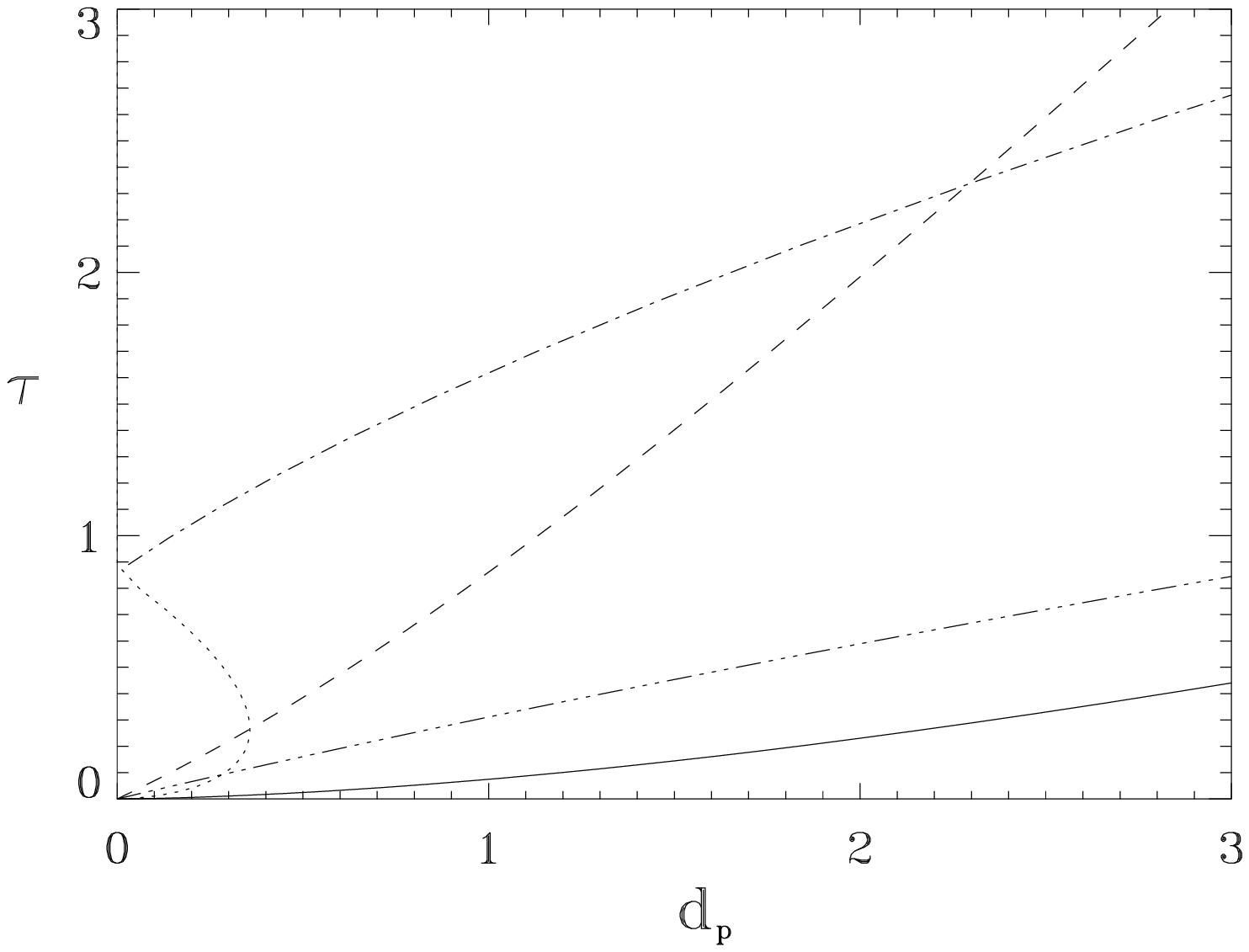}
\caption{A quintessence universe with model parameters 
$\Omega_{m0} = 0.30$, $\Omega_{Q0} = 0.70$ and  $w_Q = - 0.50$.
(a) The past light cone (dotted line) and the future light cone (dot-dashed line) 
at the present epoch in a $\eta$ -  $\chi$ diagram.
Also shown are the Hubble surface (the dashed curve), 
the particle or visual horizon (dash-triple dotted line) and the event horizon (solid line). 
Here ${{\eta}}_0 = 3.1$ and ${{\eta}}_{max} = 7.7$. 
(b) Same as a in a $\tau$ - $d_p$ diagram.
\label{fig9}
}
\end{figure}
%%%%%%%%%%%%%%%%%%%%%%%%%%%%%%%%%%%%%%%%%%%

Figure~\ref{fig9} shows the past and future light cones, the Hubble surface, the particle
horizon and the event horizon for our $Q$-model, and should 
be compared to Figure~\ref{fig2} for the standard $\Lambda$-model.
Applying the methods introduced in \S~\ref{obs} to quintessence models in general 
it is easy to see that once quintessence completely dominates the expansion, the proper 
distance to the Hubble surface is given by 
\begin{equation}
d_{hs}(\tau) \approx \frac{R_{H0}}{n}\;\tau = \frac{3 (1 + w_Q)  R_{H0}}{2}\;\tau
\end{equation}
and that the proper distance to the event horizon is
\begin{equation}
\label{accqeh}
d_{eh}(\tau) \approx \frac{2 R_{H0}}{n \;|1 + 3w_Q|}\;\tau 
                   =  \frac{3 (1 + w_Q) R_{H0}}{|1 + 3w_Q|}\;\tau \;.
\end{equation}
We emphasize that if $w_Q \geq -1/3$ there is no event horizon, and hence
the last expression (\ref{accqeh}) is only valid for models with 
$w_Q < - 1/3$. Both $d_{hs}$ and  $d_{eh}$ increase linearily
with time, but since  $d_{hs}(\tau) = |1 + 3w_Q|d_{eh}(\tau)/2 < d_{eh}(\tau)$ 
for all $- 1 < w_Q < - 1/3$,
the Hubble surface is always considerably closer than the event horizon. 
This is in contrast to models with a cosmological constant where both distances quickly approach 
the same finite limit given by equation~(\ref{hseh}).

%%%%%%%%%%%%%%%%%%%%%%%%%%%%%%%%%%%%%%%%%%
\begin{figure}[t]
\epsscale{1.}
\plottwo{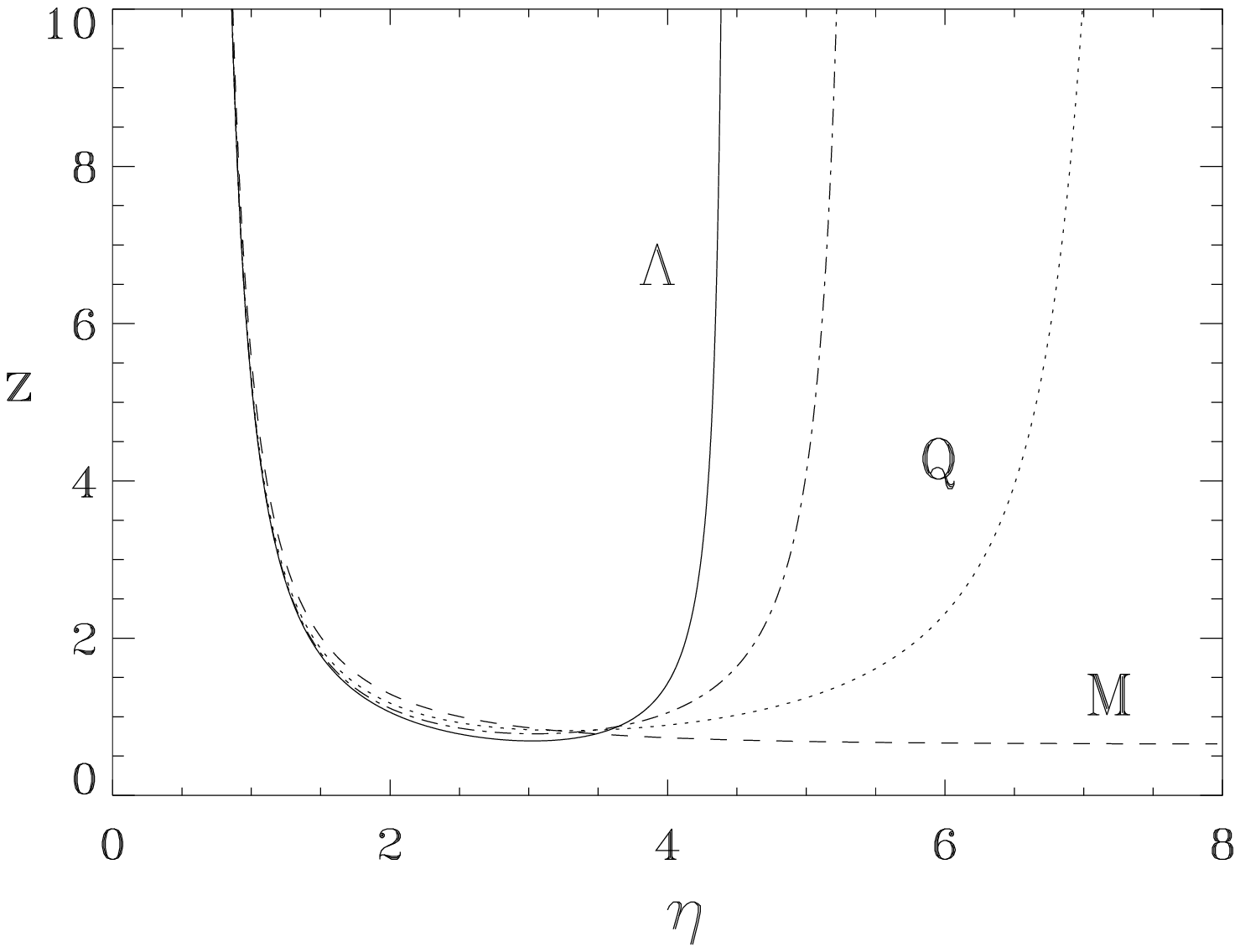}{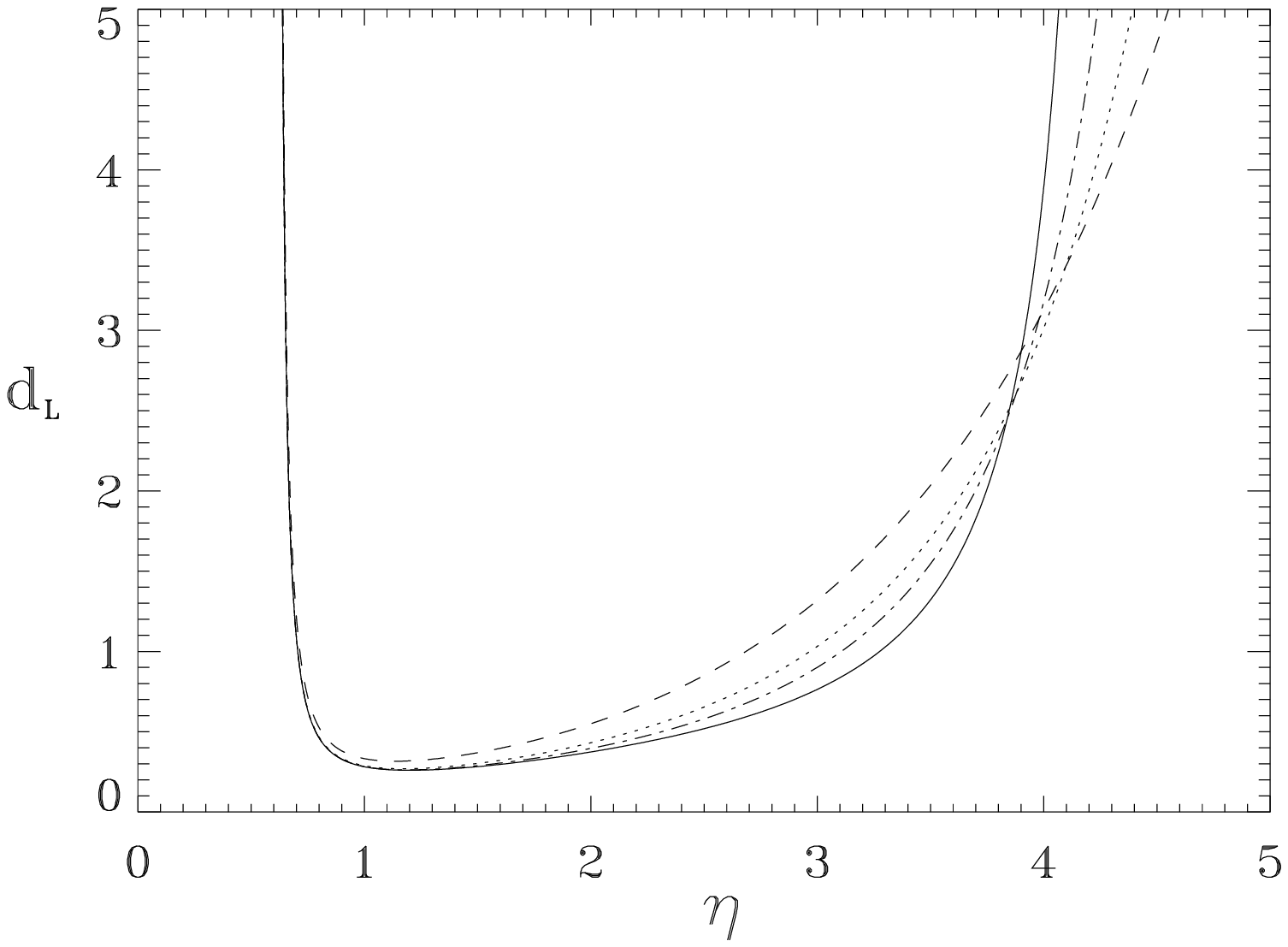}
\epsscale{2.23}
\plottwo{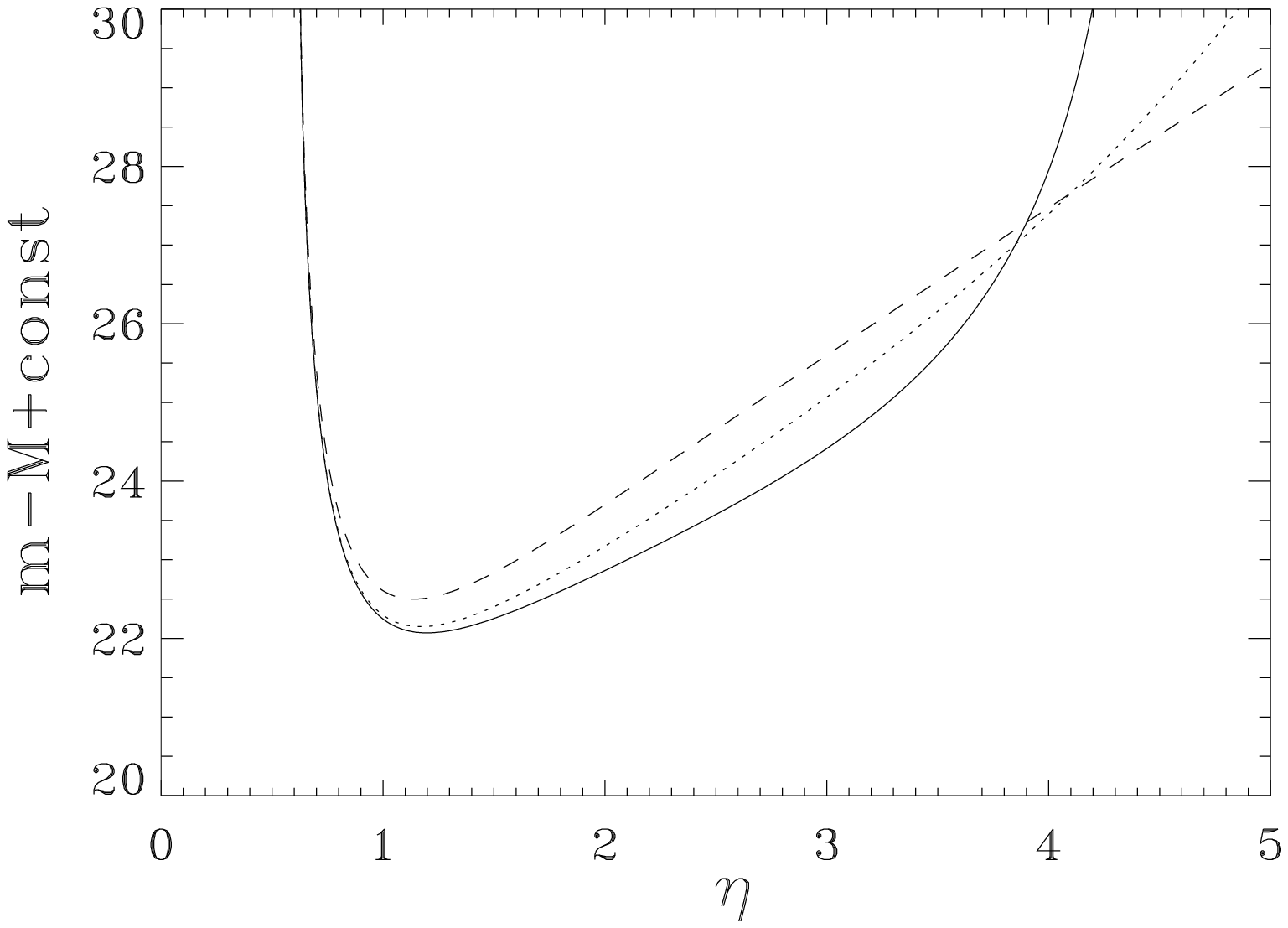}{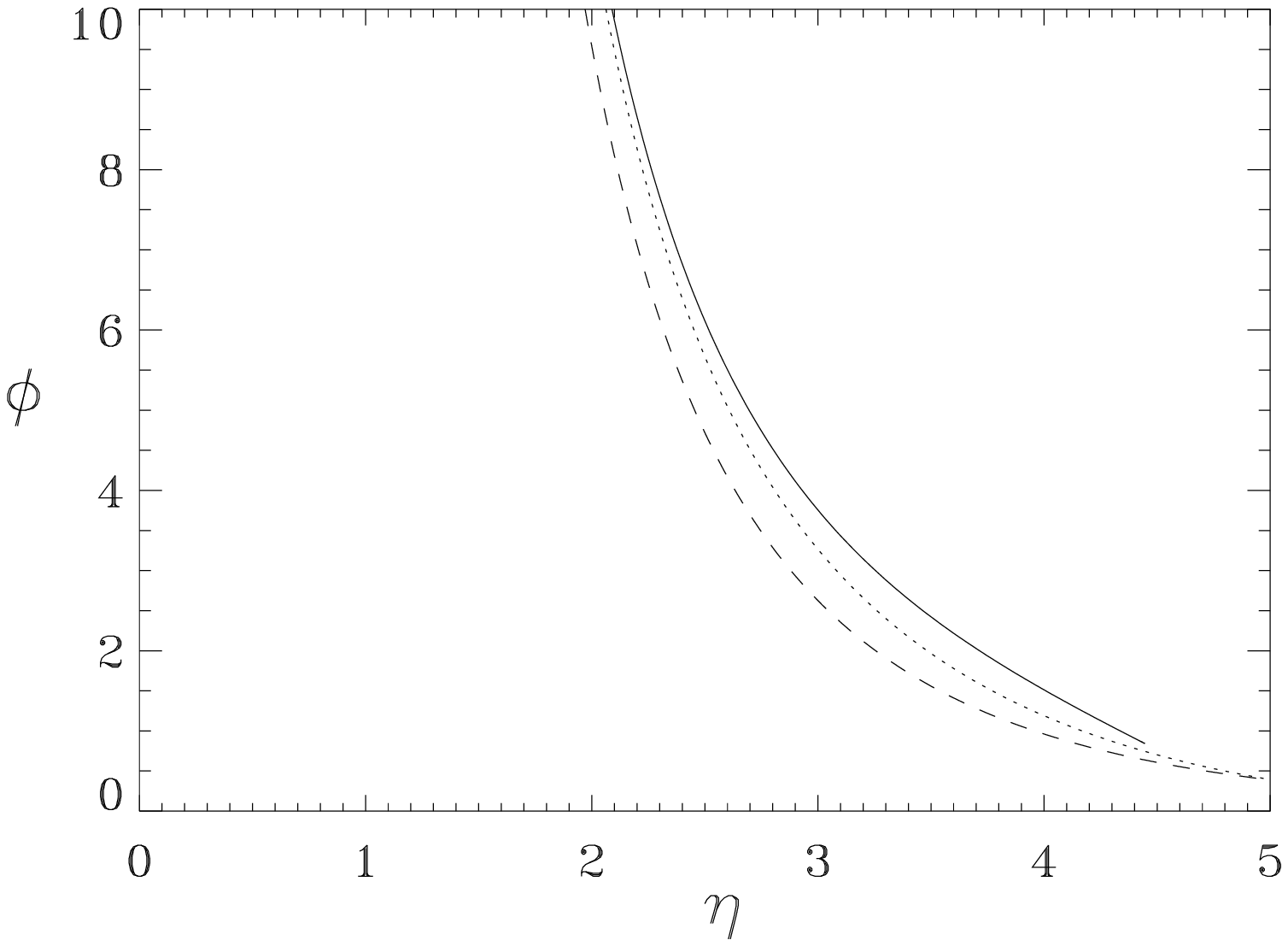}
\caption{(a) Evolution of the redshift of a source at $\chi = 0.60$ for the
model universes of Figure~\ref{fig8}: The standard $\Lambda$-model (solid curve),
the $Q$-model (dotted curve), and the $M$-model (dashed curve). Also shown is the evolution of 
the quintessence model with $w_Q = - 2/3$ (dash - dotted curve).  
(b) The luminosity distances (in units of $R_{H0}$) for the same source and models as in a.
(c) The distance moduli for the $\Lambda$, $Q$ and $M$-models.
(d)  The apparent angular sizes (in units of $D/R_{H0}$) for the  $\Lambda$, $Q$ and $M$-models.
\label{fig10} 
}
\end{figure}
%%%%%%%%%%%%%%%%%%%%%%%%%%%%%%%%%%%%%%%%%%%

In order to investigate the evolution of the redshift, the apparent magnitude and 
the angular size of distant sources in a quintessential universe we pick a source 
at a typical conformal distance $\chi = 0.60$. 
The results are shown in Figure~\ref{fig10} for the models of Figure~\ref{fig8}. 
Note that at late times
\begin{equation}
z \sim (n\; \Omega_{Q0}^{1/2}\; \tau)^{n}
\end{equation}
and
\begin{equation}
d_L \sim (n\; \Omega_{Q0}^{1/2}\; \tau)^{2 n} \;.
\end{equation}
Hence $m$ is a linear function of $\log{(\tau)}$.
Furthermore the minimum size of $\phi$ is given by equation~(\ref{accphi}) as before.
These results should be compared to the late time behaviour of $z$, $d_L$ and $m$ 
in models with a cosmological constant, shown by equations~(\ref{accz}), (\ref{bdl}) and 
(\ref{bm}) and in
Figures~\ref{fig4} and \ref{fig5}.

%--------------------------------
\subsection{The $Q$-sphere}

In \S~\ref{redbrightsize} we defined the  $\Lambda$-sphere as the surface bounding
the region in our visible universe where the expansion is accelerating due to the 
presence of the cosmological constant. For the quintessence models we similarly
define a $Q$-sphere as the surface in the visible universe
bounding the accelerating region driven by the quintessence component.
 
We denote the time when quintessence acceleration becomes dominant by $\tau_Q$ and use 
equation~(\ref{algebra}) with $\Omega_{\Lambda0} = 0$ to find the 
value $a_Q = a(\tau_Q)$. It is given by
\begin{equation}
\label{aQ}
a_Q = \left(\frac{|1 + 3w_Q| \Omega_{Q0}}{\Omega_{m0}}\right)^{1/3w_Q}\;.
\end{equation}
As a result, at any time $\tau > \tau_Q$ the observed redshift, $z_Q$, 
of a source which emitted its light at $\tau_Q$ is given by
\begin{equation}
\label{Qmin}
1 + z_Q = \frac{a (\tau)}{a_{Q}} = 
                  \left(\frac{|1 + 3 w_Q|) \Omega_{Q0}}{ \Omega_{m0}}\right)^{- 1/3w_Q}a(\tau) \;.
\end{equation}

\begin{figure}[t]
\epsscale{1.}
\plottwo{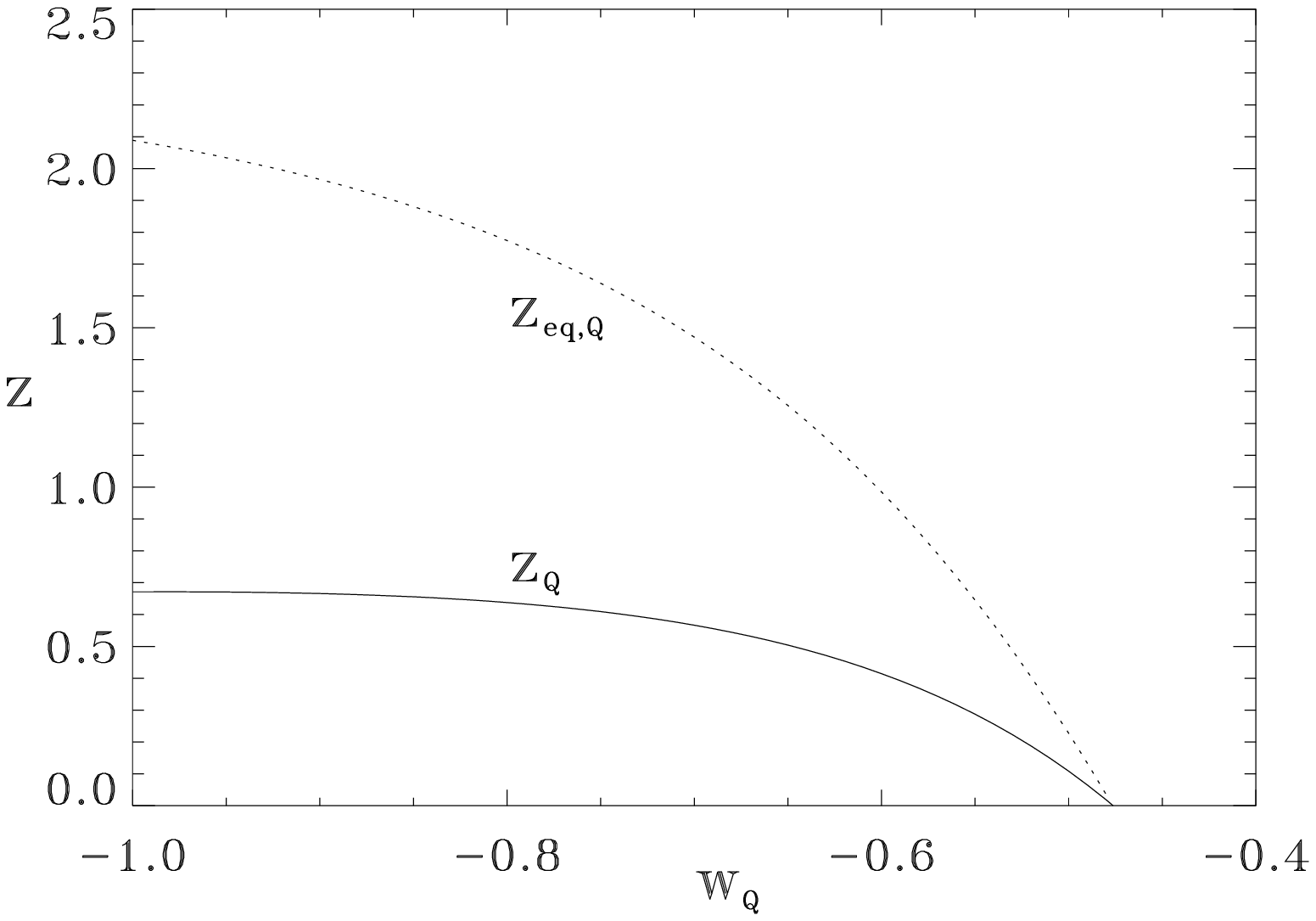}{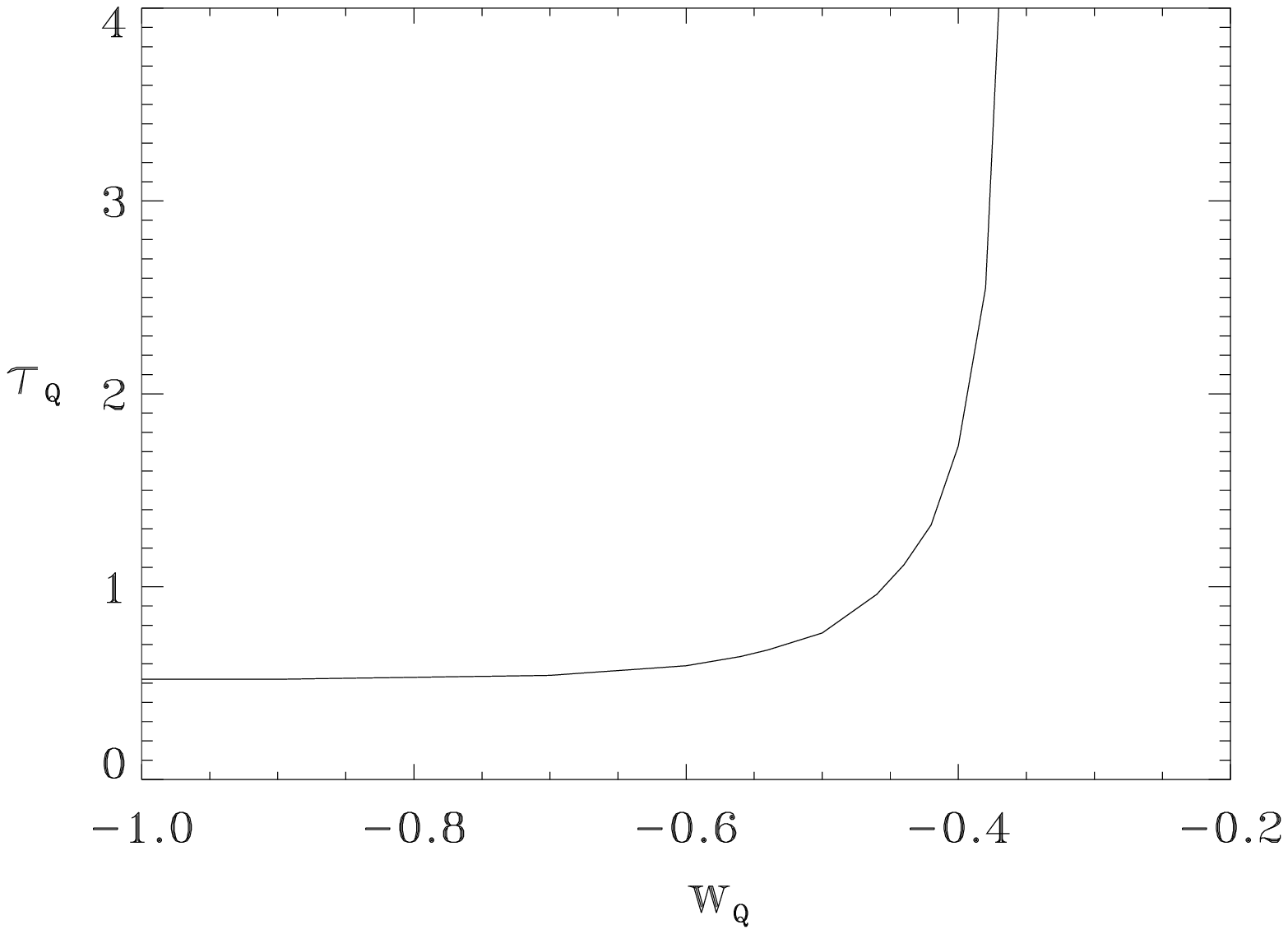}
\caption{The $Q$-sphere at the present epoch for models with $\Omega_{m0} = 0.30$ and 
$\Omega_{Q0} = 0.70$.  
(a) The redshift $z_Q$ as a function of $w_Q$. Also shown is the redshift $z_{eq,Q}$. 
(b) The time $\tau_Q$ as a function of $w_Q$. \label{fig11}
}
\end{figure}

In the same way as in the previous section we find from equations~(\ref{zder}) and (\ref{timescales}) 
that in a universe dominated by quintessence, $d (1 + z)/ d\tau =  dz/ d\tau$ 
is zero, and $T_a(\tau) = (1 + z)T_a(\tau_{em})$, 
when $(da/d\tau)_{\tau} = (da/d\tau)_{\tau_{em}}$. This particular epoch corresponds to a redshift
$z = z_{eq,Q}$ given by the solution of the algebraic equation
\begin{equation}
\label{zQzero}
\frac{\Omega_{m0}}{\Omega_{Q0}} a^{3w_Q} [(1 + z) - 1] = 1 - (1 + z)^{1 + 3w_Q} \;.
\end{equation}
where we have used equation~(\ref{mainq}). This equation can easily be solved 
numerically, and Figure~\ref{fig11} shows $z_{eq,Q}$ at the present epoch,
together with $z_Q$ and $\tau_Q$, as functions of $w_Q$ for models with $\Omega_{m0} = 0.30$ and 
$\Omega_{Q0} = 0.70$.
In general, as $w_Q$ approaches $- 1$, the redshift $z_{eq,Q}$ approaches
the value $z_{eq}$ for the $\Lambda$-model (eq.~[\ref{zzero}] with $\Omega_{Q0}$ instead of 
$\Omega_{\Lambda0}$). In the same way $z_Q$ and $\tau_Q$ approach $z_{\Lambda}$ and 
$\tau_{\Lambda}$ respectively.
For the special case $w_Q = - 2/3$ equation~(\ref{zQzero}) has a simple solution: 
$1 + z_{eq,Q} = (\Omega_{Q0} / \Omega_{m0}) a^2(\tau) = (a(\tau) / a_Q)^2$.
Note that since $1 + z \geq 1$, equations~(\ref{Qmin}) and (\ref{zQzero}) have physical
solutions only if $a \geq a_Q$. 
 
%%%%%%%%%%%%%%%%%%%%%%%%%%%%%%%%%%%%%%%%%%
%%%%%%%%%%%%%%%%%%%%%%%%%%%%%%%%%%%%%%%%%%% 

Figure~\ref{fig11} also shows that if $w_Q$ is less than $\approx - 2/3$, 
then $z_Q$ and $\tau_Q$ vary slowly
with changing $w_Q$, 
indicating that at the present epoch numerical values of observables are not very sensitive to $w_Q$ 
in the range $-1 < w_Q < - 2/3$. 
The same thing can be inferred from Figures~\ref{fig8} and~\ref{fig10}. 
This is similar to the corresponding results of \cite{gud90} 
for FRW-models without a cosmological constant, which show that for models with the same value of $q_0$
the classical cosmological tests are degenerate at low redshifts with respect to different values of 
the pressure parameter $w_i$ (see eq.~[\ref{eos}]). 
It is therefore necessary to go to high redshifts 
in order to distinguish between the various models (see also~\cite{mao01}).

\begin{figure}[t]
\epsscale{0.5}
\plotone{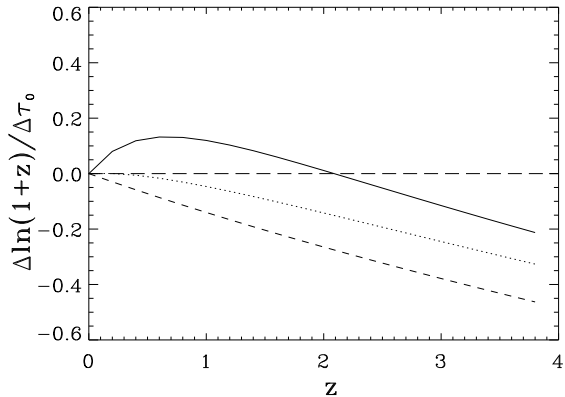}
\caption{$(\Delta (1 + z) / \Delta \tau_0)/(1 + z)$ at the present epoch 
as a function of $z$ for different model universes of Figures~\ref{fig8}:
The standard $\Lambda$-model (solid curve), the $Q$-model  (dotted curve), 
and the $M$-model (dashed curve). \label{fig12}
}
\end{figure}

Continuing with the general approach already introduced for models with a cosmological constant, 
we can similarly determine 
changes in observable quantities over extended periods of observing time.
In Figure~\ref{fig12} we show the relative change
$(\Delta (1 + z) / \Delta \tau_0)/(1 + z) = 1 / T_z$ at the present epoch
as a function of $z$ for our $Q$-model.
We find that $a_Q = 0.90$, $\tau_Q = 0.77$, $z_Q = 0.11$ and $z_{eq,Q} = 0.23$. In this
case $\tau_0 = 0.87$, so cosmic acceleration has started relatively recently, 
i.e.\ about $1.4$ Gyrs ago if $h_0 \approx 0.70$.

%%%%%%%%%%%%%%%%%%%%%%%%%%%%%%%%%%%%%%%%%%
%%%%%%%%%%%%%%%%%%%%%%%%%%%%%%%%%%%%%%%%%%%

Finally, let us investigate the extent of the observable universe in the $Q$-model 
as well as the question of causal connections. We use the same terminology and notation as 
in \S~\ref{causal} so the results can easily be  compared to the corresponding
results for the $\Lambda$-model. 
In the $Q$-model with $\Omega_{m0} = 0.30$, $\Omega_{Q0} = 0.70$ and  $w_Q = - 0.50$
we find that ${\eta}_0 = 3.1$, ${\eta}_{max} = 7.7$,
${\eta}_{em\star} = 0.72$
and $z_{\star} = 24$. This means, that the light being emitted now by
sources having redshift greater than $24$, will not reach us 
until the sources have crossed our event horizon. This particular value 
is much larger than the highest measured redshift at the present time.
 
In this model we also have that $\chi_c = 4.7$ which is further away than 
the particle horizon at $\chi_{ph} = 3.1$.
Hence, none of the  sources that we can observe at the present epoch have yet crossed
our event horizon. As a result we presently 
have causal contact with all of them.
The relative number of sources presently within the event horizon is equal to
$(\chi_c(\eta_0) / \chi_{eh}(0))^3 = 0.23$. This means that in this model
77\% of all sources initially within our 
observable part of the universe have already crossed the event horizon. 
So far we have not seen any of these departed sources but they will appear in our sky in the future and 
show themselves as they were in a younger universe.

The portion of the observable sources that we could already have seen in principle
is given by $(\chi_{ph}(\eta_0) / \chi_{eh}(0))^3 = 0.065$. Hence,
in this model, we have yet to  see about 93\% of the observable sources.

It should be emphazised that we use this $Q$-model only as a pedagogical example in order
to show the effects of $w_Q$ on the various numerical results.
The astronomical observations indicate that the high value of $w_Q$ corresponding to this model 
may not be realistic, and that a value closer to the $\Lambda$-value of 
$- 1$ is more likely to be correct. 
One should also keep in mind that observational results are not very sensitive to
the value of $w_Q$, if $- 1 < w_Q < - 2/3$.

%---------------------------- 8 ----------------------------------------
\section{DISCUSSION AND CONCLUSIONS}
\label{discussion}

In this paper we have discussed the evolution of ever-expanding Big Bang universes
which undergo acceleration, either due to a cosmological constant, $\Lambda$, or a 
quintessence field, $Q$. In particular we have investigated the evolution
of our observable part of the universe with emphasis on the evolution of our past light cone, 
the Hubble sphere and the particle and event horizons. The $\Lambda$-sphere (or $Q$-sphere,
in the case of quintessence), which is the surface bounding the region 
in our visible universe where cosmic acceleration dominates, has also been 
investigated in detail. 
We have traced observables such as  redshift, apparent magnitude and 
apparent angular size of distant sources through cosmic history, 
and shown in considerable detail how their images change and fade once the cosmic expansion
is  accelerating. 

Taking at face value recent observations which indicate that 
$\Omega_{\Lambda0} = 0.70$, $\Omega_{m0} = 0.30$ and $h_0 = 0.70$
we find that the universe is presently about $13.5$ Gyrs old, and that 
cosmic acceleration started $6.1$ Gyrs ago, 
well before the formation of the solar system. The $\Lambda$-sphere
is presently at a redshift of $0.67$, and the redshifts of sources out to a redshift of $2.1$
are increasing with time 
due to the influence of $\Lambda$. Further out on the light cone the redshift is 
decreasing as in a universe without a cosmological constant. 
Within a few Hubble times the event horizon will be stationary at a fixed proper distance of 
$5.1$ Gpc. This distance limits the extent of our observable universe for all time.

Cosmic sources with redshifts in the range $0.68$ to $1.7$ are now emitting light that 
will not reach us until these sources have crossed our event horizon. At that 
time they will be completely out of causal contact with us.
All sources with redshift larger than $1.7$ 
have already crossed the event horizon and are thus out of causal contact.

About 98\% of all sources originally within our observable part of the universe have by now
crossed the event horizon. Because of the finite speed of light we still ``see'' a large 
portion of these sources (about 40\%, the ones inside our particle horizon) 
and will eventually be able to see 
them all. They will appear as they were in the distant past before crossing the event horizon.
Because of redshift effects, all these sources will, however, fade away
and disappear from view on a timescale measured in a few Hubble times.

The quintessence models become equivalent to a model with a cosmological constant 
in the limit when $w_Q$ tends to $- 1$, and
all quintessence models with $w_Q$ in the range  $-1 < w_Q < - 1/3$ have event horizons.
For $w_Q$ less than about $- 2/3$ the numerical values of various observables at the present epoch
do not depend critically on $w_Q$, and are very similar to the values for $w_Q = -1$. Hence there is an
observational degeneracy with respect to $w_Q$ in that range.

Returning to the standard $\Lambda$-model it is clear that the future evolution of the
observable universe is rather bleak from the human point of view. The receding galaxies 
will approach the cosmic event horizon on a timescale measured in
a few Hubble times. Observers will not see the event horizon as such, 
but as galaxies approach it, their apparent motion slows down because of time dilation, 
and finally they will appear to be hovering at the horizon. This
is why the apparent angular size of galaxies tends to a finite value at
infinite time.
 
Because of the exponentially increasing redshift, the apparent luminosity of the
galaxies decreases on a timescale of a few Hubble times, making them disappear from view.
If observers had instruments sensitive enough so that
they could follow the galaxies for all time, they would eventually see all the matter 
originally within the observable universe forming a membrane at the event horizon. 
This is analogous to what an observer, stationed far away from a black hole, would
see if he was watching luminous matter falling into the black hole. 

From this discussion it is clear that in $\Lambda$-models (and also in quintessence models, 
as long as $- 1 < w_Q < - 1/3$) any fundamental observer (more precisely his local supercluster) 
will be left alone in his observable universe within a few Hubble times after the Big Bang. 
This rather dismal prospect raises interesting questions about the future evolution 
of life in the universe and cosmic communication.
We shall not tackle such questions here but instead refer the reader  to the papers by
\cite{gott96} and 
\cite{kra00} which discuss the fate of life in an accelerating universe.

Finally we mention that if the expansion is dominated by a cosmological constant one 
expects that eventually there will be upward quantum fluctuations making bubbles of high
density vacuum which consequently undergo inflation~\citep{garriga98,linde86}. Also,
if for some reason the cosmological constant 
(or the quintessence field) were to decay, the future evolution of the observable
universe would be different from the scenario presented here
(see e.g.\ \cite{stab99} and \cite{bar00} for a discussion of various possibilities).

%--------------------------------- --------------------------------------
\acknowledgments

We are grateful to L\'{a}rus Thorlacius, Thorsteinn S{\ae}mundsson and an anonymous
referee for useful comments.  
This work was partially supported by the Research Fund of the University of Iceland.

%%%%%%%%%%%%%%%%%%%%%%%%%%%%%%%%%%%%%%%%%%%%%%%%%%%%%%%%%%%%%%%%%%%%%%%%
%-----------------------------------------------------------------------

%We are grateful to N. N. for
%doing the math in section~\ref{bozomath}.
%More information on the AASTeX macros package are available at
%\url{http://www.aas.org/publications/aastex} or the
%\anchor{ftp://www.aas.org/pubs/}{AAS ftp site}.

%----------------------------------------------------------------------

%\appendix

%\section{Appendicial material}

%Consider once again

%% Appendix material should be preceded with a single \appendix command.
%% There should be a \section command for each appendix. Mark appendix
%% subsections with the same markup you use in the main body of the paper.

%% Each Appendix (indicated with \section) will be lettered A, B, C, etc.
%% The equation counter will reset when it encounters the \appendix
%% command and will number appendix equations (A1), (A2), etc.

%---------------------------------------------------------------------

\pagebreak

%%%%%%%%%%%%%%%%%%%%%%%%%%%%%%%%%%%%%%%%%%%%%%%%%%%%%%%%%%
\end{document}